\documentclass[iop]{emulateapj}
\renewcommand{\baselinestretch}{1.0}

\usepackage{graphicx}
\usepackage{amsmath}
\usepackage{natbib}
\newcommand{\pan}{{\tt Pandurata}}

\shorttitle{Dark Matter around Black Holes}
\shortauthors{Schnittman}

\begin{document}

\title{The Distribution and Annihilation of Dark Matter Around Black Holes}

\author{Jeremy D.\ Schnittman}
\affil{NASA Goddard Space Flight Center, Greenbelt, MD 20771}
\affil{Joint Space-Science Institute, College Park, MD 20742}
\email{jeremy.schnittman@nasa.gov}

\begin{abstract}
We use a Monte Carlo code to calculate the geodesic orbits of test
particles around Kerr black holes, generating a distribution function
of both bound and unbound populations of dark matter particles. From
this distribution function, we calculate annihilation rates and
observable gamma-ray spectra for a few simple dark matter models. The
features of these spectra are sensitive to the black hole spin,
observer inclination, and detailed properties of the dark matter
annihilation cross section and density profile. Confirming earlier
analytic work, we find that for rapidly spinning black holes, the
collisional Penrose process can reach efficiencies exceeding $600\%$,
leading to a high-energy tail in the annihilation spectrum. 
The high particle density and large proper volume of the
region immediately surrounding the horizon ensures that the observed
flux from these extreme events is non-negligible.
\end{abstract}


\section{INTRODUCTION}\label{section:intro}

Prompted by the recent paper by \citet{Banados2009} [BSW], there has been a
great deal of interest in the potential of Kerr black holes
to accelerate particles to ultra-relativistic energies and thus to
probe a regime of physics otherwise inaccessible. The vast majority of
this work has been analytic and thus largely limited to the most
simple photon and particle trajectories in the equatorial plane. Here
we present a more numerical approach that focuses on calculating the
fully relativistic distribution function of massive test particles
around a spinning black
hole. With this distribution function and a simple model for the dark
matter annihilation mechanism, we can then calculate the
annihilation rate and observed spectrum as a function of black hole
spin and observer inclination. 

It has been noted repeatedly in recent works that the net energy
gained through the Penrose process is quite modest, as is the
fraction of collision products that might escape, and thus the
astrophysical importance of the BSW effect is questionable
\citep{Jacobson2010,Banados2011,Harada2012,Bejger2012,McWilliams2013}. We argue
here that two primary factors (to our knowledge largely neglected in
previous work) could greatly enhance the astrophysical relevance and
observability of this annihilation. The first is an energy-dependent
cross section for dark matter (DM) annihilation. This could take many
forms, the simplest of which are p-wave annihilation
\citep{Bertone2005,Chen2013,Ferrer2013}, where the 
cross section scales like the relative velocity, or a threshold
energy, above which the cross section increases greatly. This latter
assumption is a natural choice for a model that includes multiple DM
species, with the more massive particles intermediate products
in the annihilation process towards gamma rays [see, e.g.,
\citet{Zurek2014}]. Because gravity is the
only known force capable of accelerating dark matter particles to
high energies, it is possible that new annihilation channels could
occur around black holes that are completely inaccessible everywhere
else in the universe.

The other effect considered here is the relativistic enhancement of
the density close to the black hole. This is due to the time dilation of
observers near the horizon. In a steady-state system, one can think of
dropping particles into the black hole from infinity at a constant
rate $\Gamma_\infty$ as measured by coordinate 
time $t$. To an observer near the black hole measuring proper time
$\tau$, an enhanced rate $\Gamma_\infty (dt/d\tau)$ is seen, with
$dt/d\tau>1$. For annihilation rates that scale like the density
squared, the local annihilation rate will be enhanced by
$(dt/d\tau)^2$. Of course, the products will get redshifted on their
way back out to an observer at infinity \citep{McWilliams2013}, but we
are still left with a net enhancement of $dt/d\tau$. 

Even without this relativistic enhancement, numerous models also
predict an astrophysical enhancement of the dark matter density in
the galactic nucleus. Adiabatic growth of the central black hole
will capture a large number of particles onto tightly bound orbits,
growing a steep density spike as the black hole grows
\citep{Gondolo1999,Sadeghian2013}. Gravitational scattering off the
dense nuclear star cluster will also lead to a dark matter spike
\citep{Gnedin2004}, similar to the classical two-body scattering
result of \citet{Bahcall1976}. At the same time, self-annihilation
\citep{Gondolo1999} and elastic scattering \citep{Shapiro2014,Fields2014}
will act to flatten out this spike into a shallow core more similar
to the unbound population.

Because our approach to this problem is predominantly numerical, we
can easily include treatment of a range of black hole spins, particle
distributions, and cross sections, and not limit ourselves to special
cases with analytic solutions. Therefore, we can calculate how often
those extreme cases are likely to occur in a real astrophysical
setting [for a notable exception to the analytic approaches of
  earlier work, see the exhaustive Monte Carlo calculations of
  \citet{Williams1995,Williams2004} that explored the limits of the
  Penrose process in the context of Compton scattering and pair
  production in accretion disks and jets].
Of particular interest has been the following question: for two
particles each of mass $m_\chi$ falling from rest at infinity and
colliding near the black hole, what is the maximum 
achievable energy for an escaping photon? We find that this
limit exceeds $12 m_\chi$ for an extremal black hole with $a/M=1$,
significantly higher than previously published values of
$2.6m_\chi$ \citep{Bejger2012}. We explain the underlying reason for this
discrepancy in a companion paper \citep{Schnittman2014}.

\section{POPULATING THE DISTRIBUTION FUNCTION}\label{section:df}
\subsection{Initial conditions}\label{section:initial}
The primary goal of this paper is to calculate the 8-dimensional
phase-space distribution function $df(\mathbf{x},\mathbf{p})$ of DM
particles around a Kerr black hole. Two of these
dimensions are immediately removed due to the assumption of a
steady-state solution and stationarity of the
metric, and the mass-shell constraint of the particle momentum,
leaving us with $df(r,\theta,\phi,p_r,p_\theta,p_\phi)$. This function
is further reduced to five dimensions because axisymmetry removes the
dependence on $\phi$. 

To calculate the distribution function, we first distinguish between
two basic populations: the particles gravitationally bound and unbound
to the black hole. The properties of the bound populations are more
sensitive to underlying astrophysical assumptions, and will be
discussed below in Section \ref{section:bound}. The unbound population
is more straightforward: we simply assume an isotropic,
thermal distribution of velocities at a large distance from the black
hole. Here, ``large distance'' is taken to be the influence radius
$r_{\rm infl}$ of a supermassive black hole with mass $M$,
and the DM velocity dispersion is set equal to the stellar velocity
dispersion $\sigma_0$ of the bulge (thus the ``unbound''
population considered in this paper is still gravitationally bound
to the galaxy, just not the black hole). From the ``M-sigma'' relation
\citep{Ferrarese2000} we take 
\begin{equation}\label{eqn:M_sigma}
M \approx 2\times 10^7 M_\odot \left(\frac{\sigma_0}{100\mbox{ km/s}}\right)^4
\end{equation}
and
\begin{equation}\label{eqn:r_infl}
r_{\rm infl} \equiv \frac{GM}{\sigma_0^2} \approx 8 \mbox{ pc}
\left(\frac{\sigma_0}{100\mbox{ km/s}}\right)^2 \, . 
\end{equation}
In units of gravitational radii $r_g = GM/c^2$, the influence radius
is typically quite large: $r_{\rm infl} \approx 10^7 M_7^{-1/2} r_g$,
where $M_7 \equiv (M/10^7 M_\odot)$. 

\begin{figure*}[ht]
\caption{\label{fig:phase_space_schem} Schematic of our method for
  populating phase space with geodesic trajectories. The test
  particles are injected at large radius ($r_0= 10^7 r_g$) with
  thermal velocities with dispersion $\sigma_0 \ll c$. Those particles
  passing within $1000\, r_g$ of the black hole contribute to the
  tabulated distribution function in each volume element 
  $(r_i,\theta_j)$ through which they pass, with a weight proportional
  to the amount of coordinate time $t$ spent in that zone.}
\begin{center}
\includegraphics[width=\textwidth]{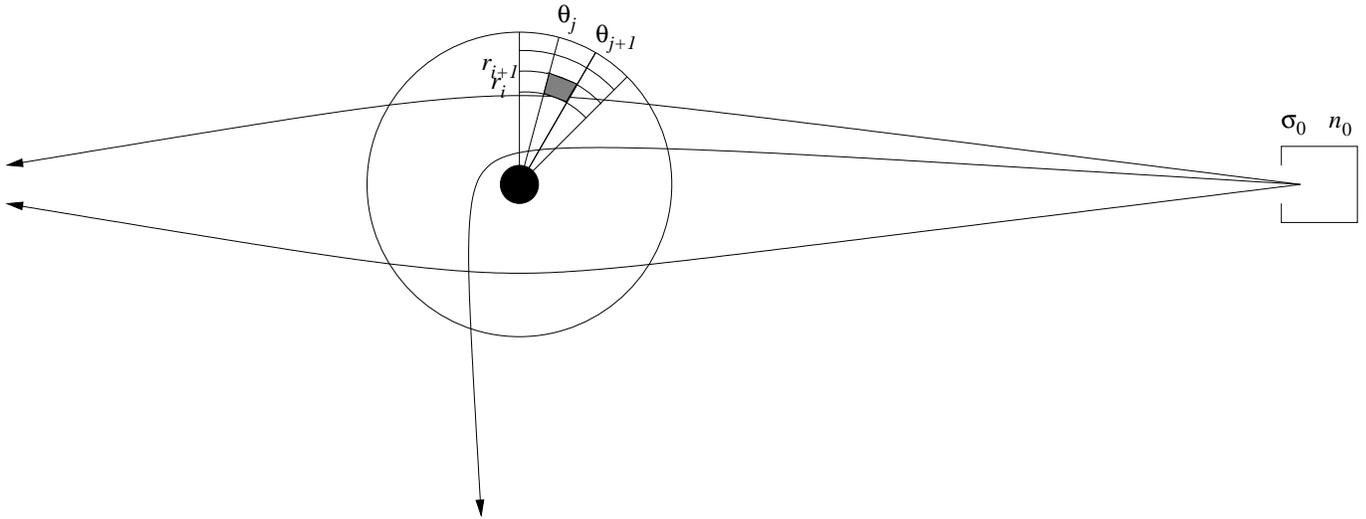}
\end{center}
\end{figure*}

Given this outer boundary condition, we shoot test particles towards
the black hole with initial velocities drawn from an isotropic
thermal distribution with characteristic velocity $\sigma_0$. As we are only
interested in the distribution function relatively close to the black
hole, we can ignore any particle with impact parameter greater than
$\approx 1000\, r_g$. For those particles that we do follow, we calculate
their geodesic trajectories with the Hamiltonian approach described in
detail in \citet{Schnittman2013} and used in the radiation transport
code \pan. A schematic of this procedure is shown in Figure
\ref{fig:phase_space_schem}. As the particle moves around the black
hole and passes through different
finite volume elements, the discretized distribution function
$df(r_i,\theta_j,\mathbf{p})$ is updated with appropriate weights. 

The great advantage of this Hamiltonian approach is that the
integration variable is the coordinate time $t$ in Boyer-Lindquist
coordinates \citep{Boyer1967}. Because this is the time measured by an observer at
infinity, it determines the rate at which particles are injected into
the system in the steady-state limit. Then the distribution function
can be populated numerically by assigning a weight to each bin in
phase space through which the test particle passes, with the weight
proportional to the amount of time $t$ spent in that volume. The
process is repeated for many Monte Carlo test particles until the
5-dimensional distribution function is completely populated. 

\subsection{Geodesics and Tetrads}\label{section:tetrads}
Following \citet{Schnittman2013}, we define local orthonormal observer
frames, or tetrads, at each point in the computational
volume. Depending on the population in question (i.e., bound
vs. unbound), it is convenient to use either the zero-angular-momentum
observer (ZAMO; \citet{Bardeen1972}) or the ``free-falling from
infinity observer'' (FFIO) tetrads. In all cases we use
Boyer-Lindquist coordinates \citep{Boyer1967}, where the metric can be
written
\begin{equation}
g_{\mu \nu} = \begin{pmatrix}
-\alpha^2+\omega^2 \varpi^2 & 0 & 0 & -\omega\varpi^2 \\
0 & \rho^2/\Delta & 0 & 0 \\
0 & 0 & \rho^2 & 0 \\
-\omega\varpi^2 & 0 & 0 & \varpi^2 \end{pmatrix}  \, .
\end{equation}

This allows for a relatively simple form for the inverse metric:
\begin{equation}
g^{\mu \nu} = \begin{pmatrix}
-1/\alpha^2 & 0 & 0 & -\omega/\alpha^2 \\
0 & \Delta/\rho^2 & 0 & 0 \\
0 & 0 & 1/\rho^2 & 0 \\
-\omega/\alpha^2 & 0 & 0 & 1/\varpi^2-\omega^2/\alpha^2 \end{pmatrix}  \, ,
\end{equation}
with the following definitions:
\begin{subequations}
\begin{eqnarray}\label{eqn:BL_equations_a}
\rho^2 & \equiv & r^2+a^2\cos^2\theta \\ 
\Delta & \equiv & r^2-2Mr+a^2 \\
\alpha^2 & \equiv & \frac{\rho^2\Delta}{\rho^2\Delta+2Mr(a^2+r^2)} \\
\omega & \equiv & \frac{2Mra}{\rho^2\Delta+2Mr(a^2+r^2)} \\
\varpi^2 & \equiv &
\left[\frac{\rho^2\Delta+2Mr(a^2+r^2)}{\rho^2}\right]\sin^2\theta \label{eqn:BL_equations_e}\, .
\end{eqnarray}
\end{subequations}
Unless explicitly included, we adopt units with $G=c=1$, so distances
and times are often scaled by the black hole mass $M$. 

The ZAMO tetrad can be constructed by 
\begin{subequations}
\begin{eqnarray}
\mathbf{e}_{(\tilde{t})} &=& \frac{1}{\alpha}\mathbf{e}_{(t)} + 
\frac{\omega}{\alpha}\mathbf{e}_{(\phi)} \\
\mathbf{e}_{(\tilde{r})} &=& \sqrt{\frac{\Delta}{\rho^2}}\mathbf{e}_{(r)} \\
\mathbf{e}_{(\tilde{\theta})} &=& \sqrt{\frac{1}{\rho^2}}\mathbf{e}_{(\theta)} \\
\mathbf{e}_{(\tilde{\phi})} &=&
\sqrt{\frac{1}{\varphi^2}}\mathbf{e}_{(\phi)} \, ,
\end{eqnarray}
\end{subequations}
where we designate tetrad basis
vectors by $\tilde{\mu}$ indices, while coordinate bases have normal
indices. 

To construct the FFIO tetrad, the time-like basis vector
$\mathbf{e}_{(\tilde{t})}$ is given by the 4-velocity
$u^\mu=g^{\mu\nu}u_\nu$ corresponding to a geodesic with $u_t=-1$,
$u_\theta=u_\phi=0$, and from normalization constraints, 
\begin{equation}
u_r = -\left[(\alpha^{-2}-1) \frac{\rho^2}{\Delta}\right]^{1/2}\, .
\end{equation}
Then the spatial basis vectors $\mathbf{e}_{(\tilde{i})}$ are
constructed via a standard Gram-Schmidt method and aligned roughly
parallel to the Boyer-Lindquist coordinate bases. 

Any vector can be represented by its components in different
tetrads via the relation
\begin{equation}
\mathbf{u} = \mathbf{e}_{(\mu)}u^\mu =
\mathbf{e}_{(\tilde{\mu})}u^{\tilde{\mu}} \, ,
\end{equation}
whereby the components are related by a linear transformation
$E_{\tilde{\mu}}^{\mu}$: 
\begin{subequations}
\begin{eqnarray}
u^\mu &=&  E_{\tilde{\mu}}^\mu u^{\tilde{\mu}} \, , \\
u^{\tilde{\mu}} &=&  [E^{-1}]_\mu^{\tilde{\mu}} u^{\mu} \, .
\end{eqnarray}
\end{subequations}
These $u^{\tilde{\mu}}$ are the components
that we use for the tabulated distribution function.\footnote{While
  these contravariant indices technically refer to 4-velocities, and
  not 4-momenta, we use the terms interchangeably in the locally flat
  tetrad basis, where most of our calculations take place.} Because of the
normalization constraints, we need only store three components of the
4-momentum in each spatial volume element, making the total
dimensionality of the distribution function five: two space and
three momentum. 

In \pan, the geodesics are integrated with a variable time step
5$^{th}$ order Cash-Karp algorithm \citep{Schnittman2013}. This
technique very naturally matches small time steps to 
regions of high curvature and thus areas of high resolution in the
spatial grid. For each time step, a weight proportional to
the coordinate time spent on that step is added to the distribution
function for that particular volume of phase space. Because 
the particle typically remains within a single volume element for many
time steps, we find that interpolation errors are small. 

The spatial momentum components
$\gamma\beta^{\tilde{i}}$ can be positive or negative and span many
orders of magnitude. To adequately resolve the phase space and capture
the relativistic effects immediately outside the black hole
horizon, we find that on order $\sim 10^3$ bins are required in
each dimension. If the entire phase-space volume were occupied, this
would correspond to an unfeasible quantity of data. Fortunately, this
volume is not evenly filled, so such a hypothetical 5-dimensional
array is in fact exceedingly sparse. In practice, we are able to use a
dynamic memory allocation technique that only stores the non-zero
elements of the distribution function. Yet even so, a well-resolved
calculation can easily require multiple GB of data for a single
distribution function, and to adequately sample this phase space
requires on the order of $\sim 10^9$ test particles, with each
geodesic sampled over thousands of time steps. Fortunately, this is a
trivially parallelizable problem, so it is relatively simple to
achieve sufficient resolution in a reasonable amount of time with a
small computer cluster. 

\subsection{Unbound Particles}\label{section:unbound}

As mentioned above, for the unbound population, the outer boundary
condition for the phase space density at $r_{\rm infl}$ is
relatively well-understood. The
velocity distribution is thermal with characteristic speed
$\sigma_0$\footnote{While there could be some small anisotropy in the
  dark matter velocity distribution at $r_{\rm infl}$, it is
  unlikely to be correlated with the black hole spin. Thus the
  predominantly radial velocities of incoming particles will be
  independent of polar angle, and therefore for all intents and
  purposes appear isotropic from the black hole's point of
  view. Similarly, even if the DM velocity distribution at the
  influence radius is not strictly Maxwellian,
  this too will have little impact on the results presented here,
  because the initial velocities are so small compared to the orbital
  velocities near the black hole, the trajectories are indistinguishable
  from particles injected from infinity with zero velocity.}. The
spatial density of dark matter is measured from galactic 
rotation curves at kpc distances from the nucleus, and then must be
extrapolated in to pc distances with a combination of observations and
stellar profile modeling. For example, in the Milky Way the DM density
near the Sun is 0.3 GeV/cm$^3$, and the radial profile can be
reasonably well-modeled with a simple $\rho \sim R^{-1}$ profile,
giving a density of $\sim 10^3$ GeV/cm$^3$ at $r_{\rm infl}$. 
Inside of $r_{\rm infl}$ there is almost certainly an additional bound
component to the DM distribution \citep{Gondolo1999}, so the
unbound population described here can best be understood as a strict
lower bound on the phase space density.

Outside of $\sim 100\, r_g$ the unbound population can be treated as a
collisionless gas of accreting particles, as in
\citet{Zeldovich1971}. In the Newtonian limit, the density and
velocity dispersion can be written
\begin{equation}\label{eqn:n_r}
n(r) = n_0 \left(1+\frac{2GM}{\sigma_0^2 r}\right)^{1/2}
\end{equation}
and
\begin{equation}\label{eqn:sigma2_r}
\sigma^2(r) = \sigma_0^2\left(1+\frac{2GM}{\sigma_0^2 r}\right)\, .
\end{equation}

\begin{figure}[h]
\caption{\label{fig:dn_dr} Spatial density ({\it a}) and mean relative
  momentum $\langle\gamma\beta\rangle_{\rm rel}$ ({\it b}) of unbound particles as
  measured in the FFIO frame in the equatorial plane of a Kerr black
  hole with $a/M=1$. The dashed lines are the
  Newtonian solutions of equations (\ref{eqn:n_r},
  \ref{eqn:sigma2_r}), while the solid curves come from the fully relativistic
  Monte Carlo calculation.}
\begin{center}
\scalebox{0.45}{\includegraphics*{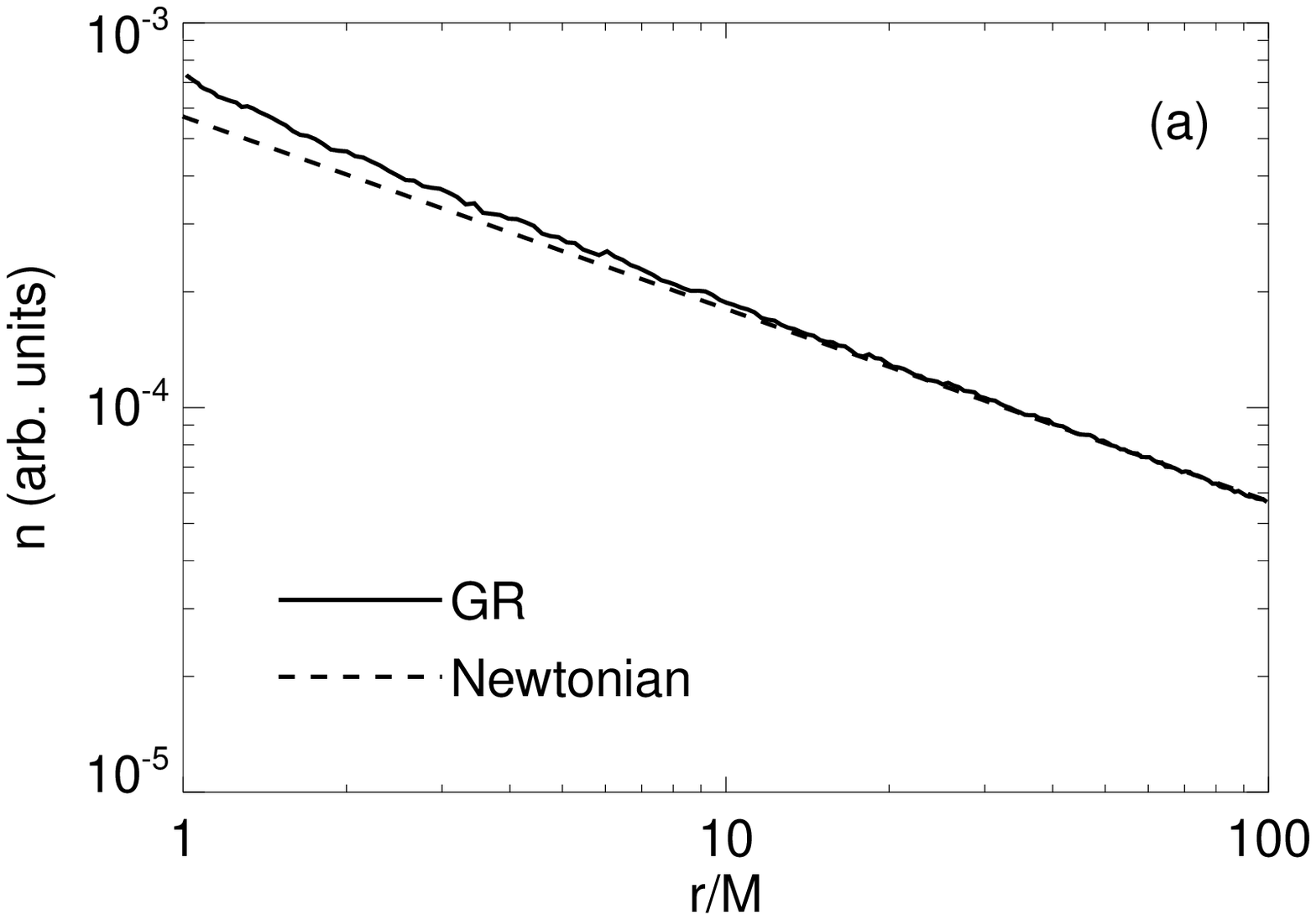}}
\scalebox{0.45}{\includegraphics*{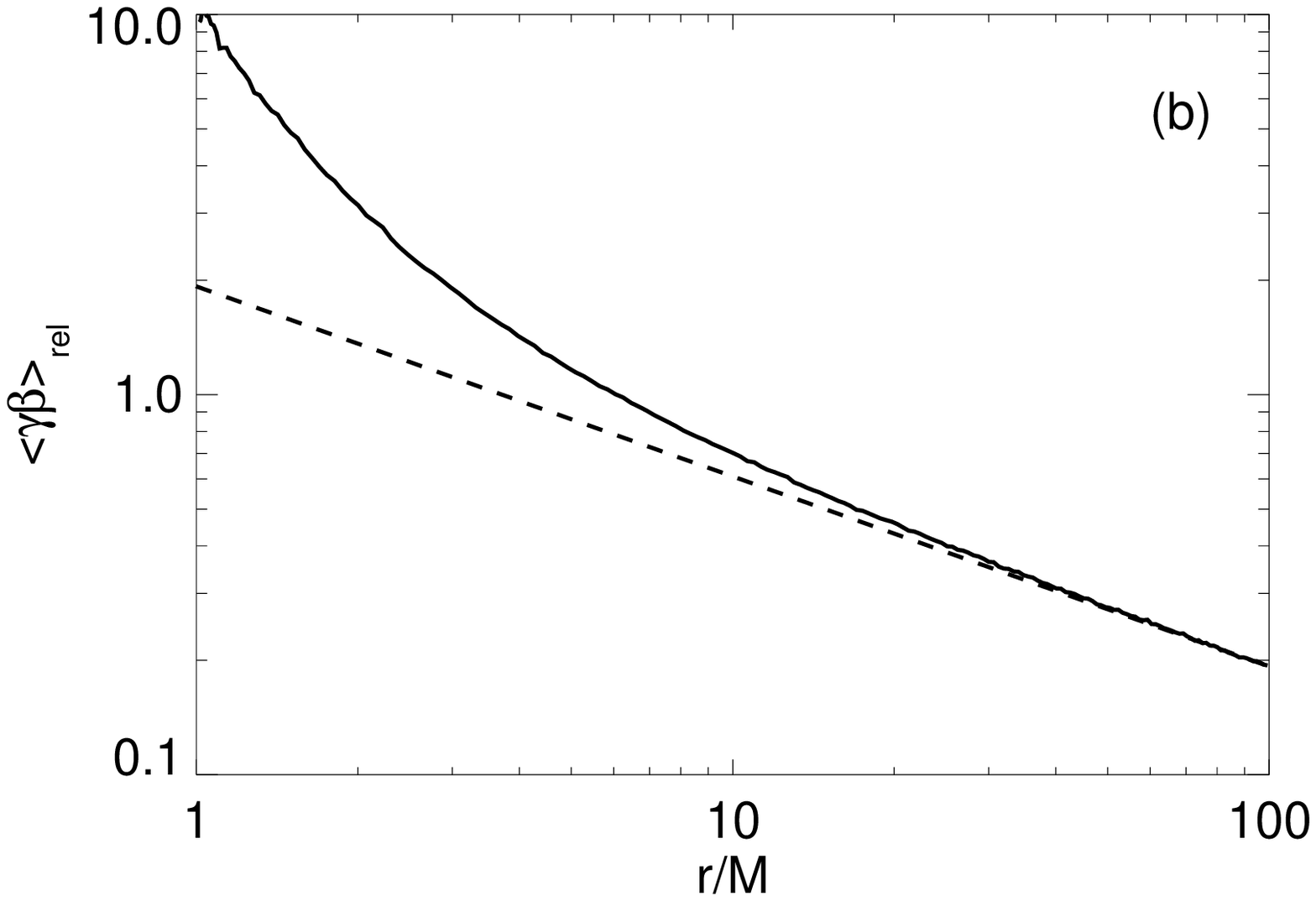}}
\end{center}
\end{figure}

In Figure \ref{fig:dn_dr} we show the spatial density of unbound
particles as measured by a FFIO around a Kerr black hole with spin
parameter $a/M=1$, as well as the mean particle momentum
as measured in that frame. 
We find very close agreement to the Newtonian results all the way
down to $r\sim 10r_g$. The deviation of the momentum from the
Newtonian solution is due largely to the special relativistic terms
proportional to the Lorentz boost $\gamma$. 

The proper density is governed by two competing relativistic
effects. One is time dilation and the other is spatial
curvature. Close to the black hole, the particle's proper time $\tau$
slows down relative to the coordinate time $t$ measured by an observer
at infinity, giving a large $dt/d\tau$. This has the effect of
increasing the number density because, in a steady state, particles
are injected into the system at a constant rate---as
measured by an observer at infinity. The injection rate measured by an observer
close to the black hole is higher by a factor of $dt/d\tau$, leading
to her seeing a larger proper density. 

\begin{figure}[h]
\caption{\label{fig:dV_dr} Proper volume measured in the FFIO
  frame. The dashed line is the Newtonian value $dV/dr=4\pi r^2$, and
  the solid curve measures the FFIO's proper volume $d\tilde{V}/dr$.}
\begin{center}
\scalebox{0.45}{\includegraphics*{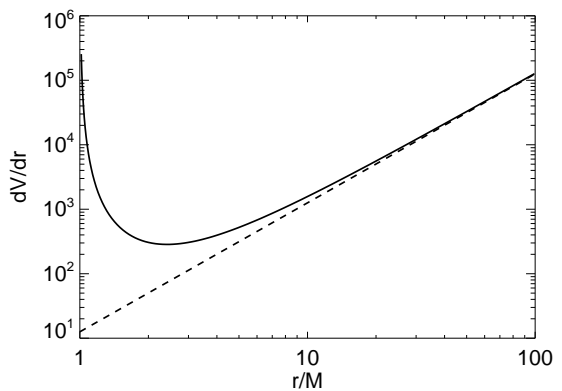}}
\end{center}
\end{figure}

In fact, the proper density would be even higher if it weren't for
another important relativistic effect: the stretching of space around a black
hole. Specifically, the Boyer-Lindquist radial coordinate element $dr$
corresponds to a greater and greater proper distance as the observer
approaches the horizon. This naturally gives a greater proper volume
$d\tilde{V}$, shown as a solid curve in Figure \ref{fig:dV_dr}. Again, we show
the Newtonian value $dV/dr=4\pi r^2$ as a dashed curve. Because the
particle interaction rates scale like $n^2\, v\, d\tilde{V}$, all these
effects combine to increase the importance of reactions near the black
hole. 

\begin{figure*}[ht]
\caption{\label{fig:df_100} Momentum distribution of unbound
  particles observed by a FFIO in the equatorial plane at radius
  $r=100M$. All particles have nearly unitary specific energy at
  infinity, so the average particle speed is very close to
  $\sqrt{2GM/r}= \sqrt{0.02}c$ ({\it panel a}). In panels ({\it
  b-d}) we show the distribution of the individual momentum
  components, which are nearly isotropic this far from the black
  hole.} 
\begin{center}
\scalebox{0.45}{\includegraphics*{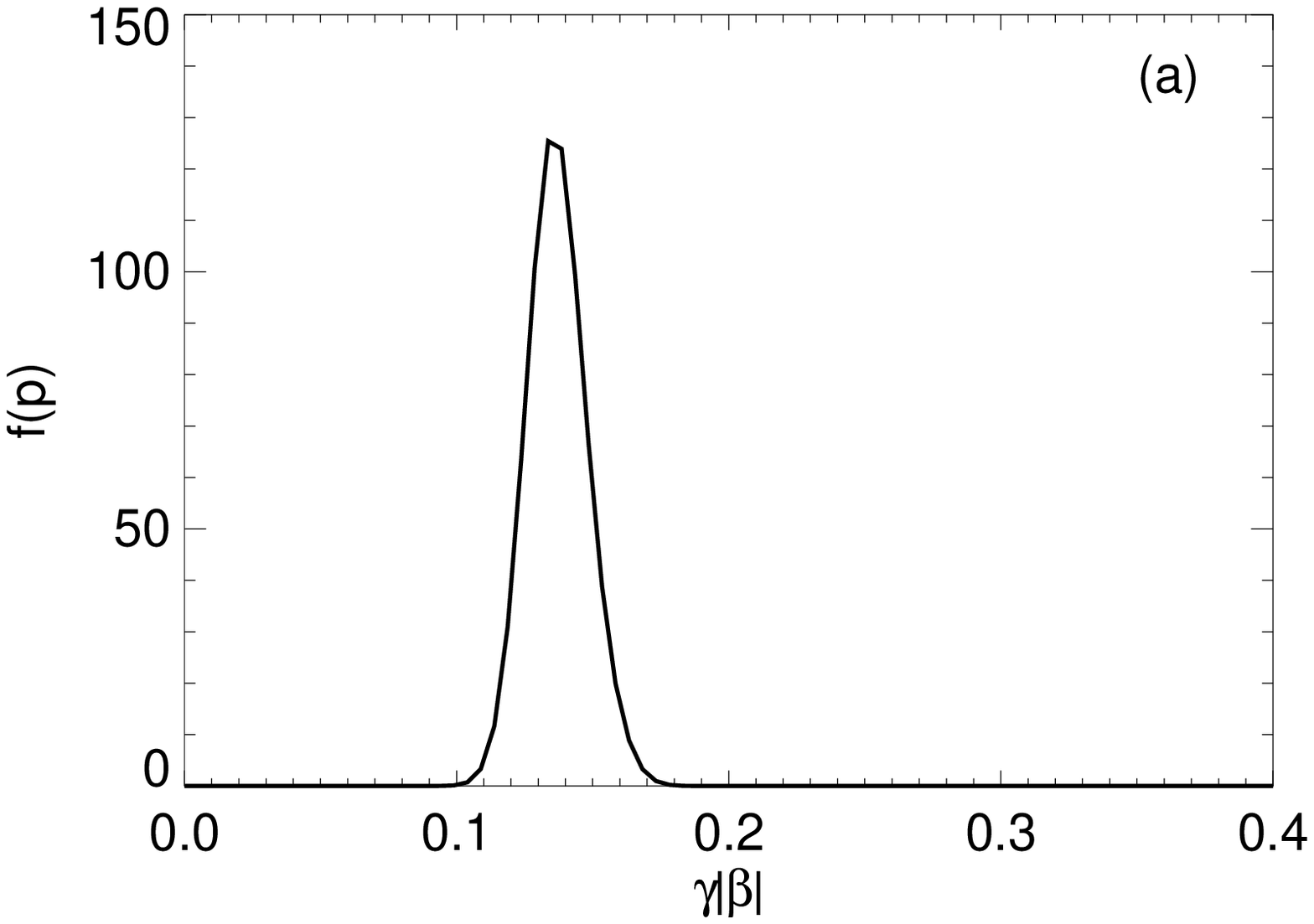}}
\scalebox{0.45}{\includegraphics*{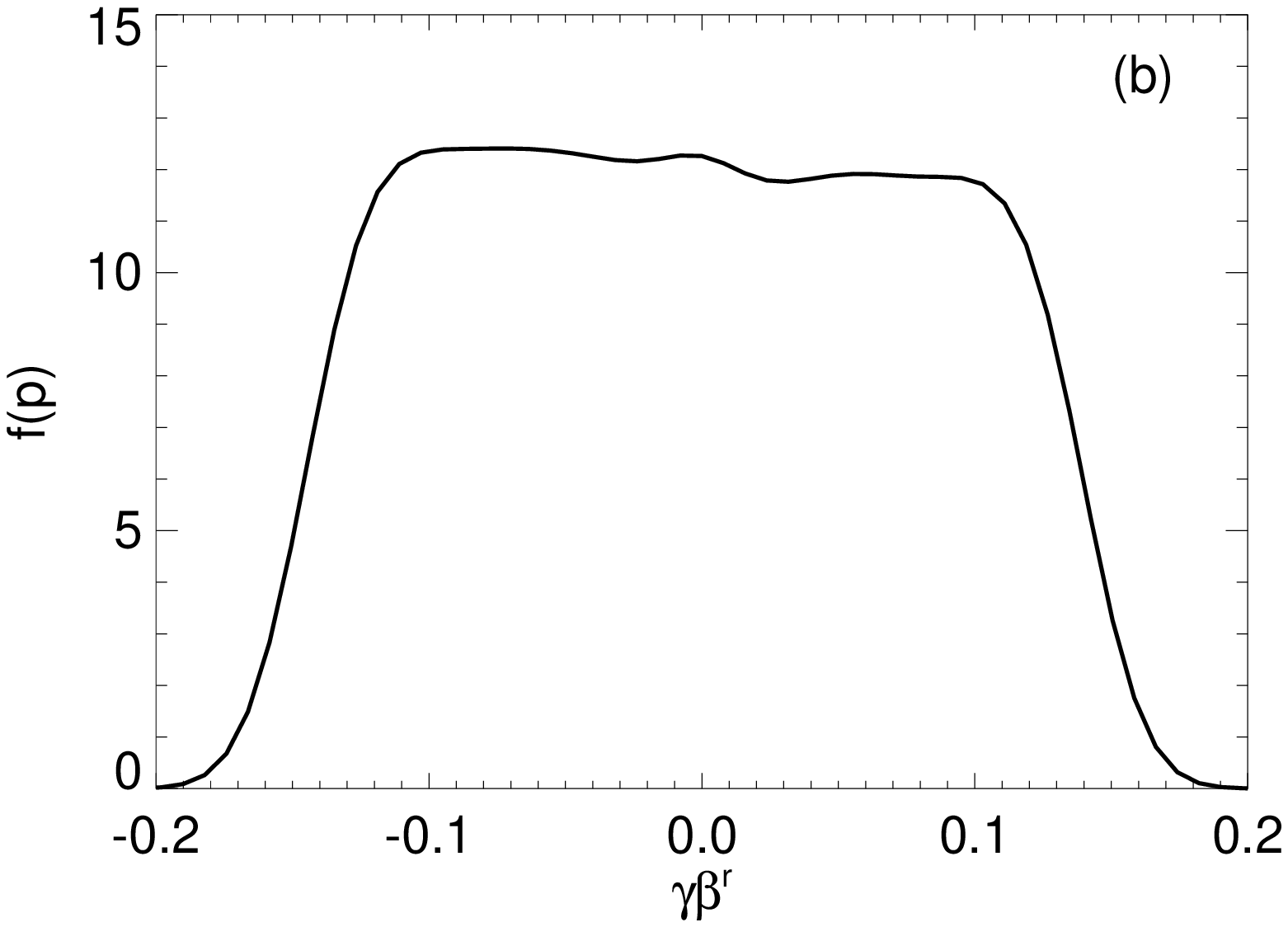}} \\
\scalebox{0.45}{\includegraphics*{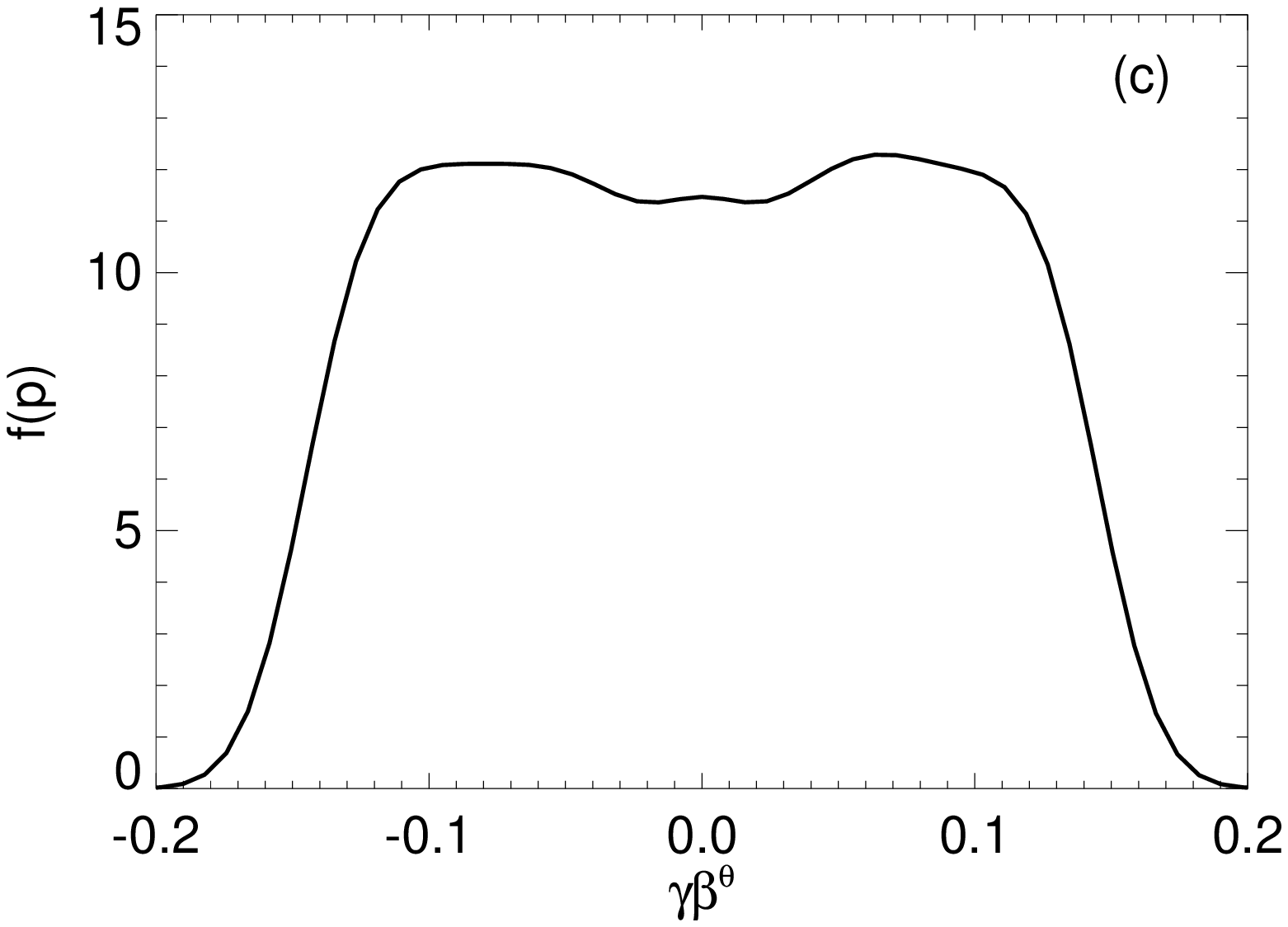}}
\scalebox{0.45}{\includegraphics*{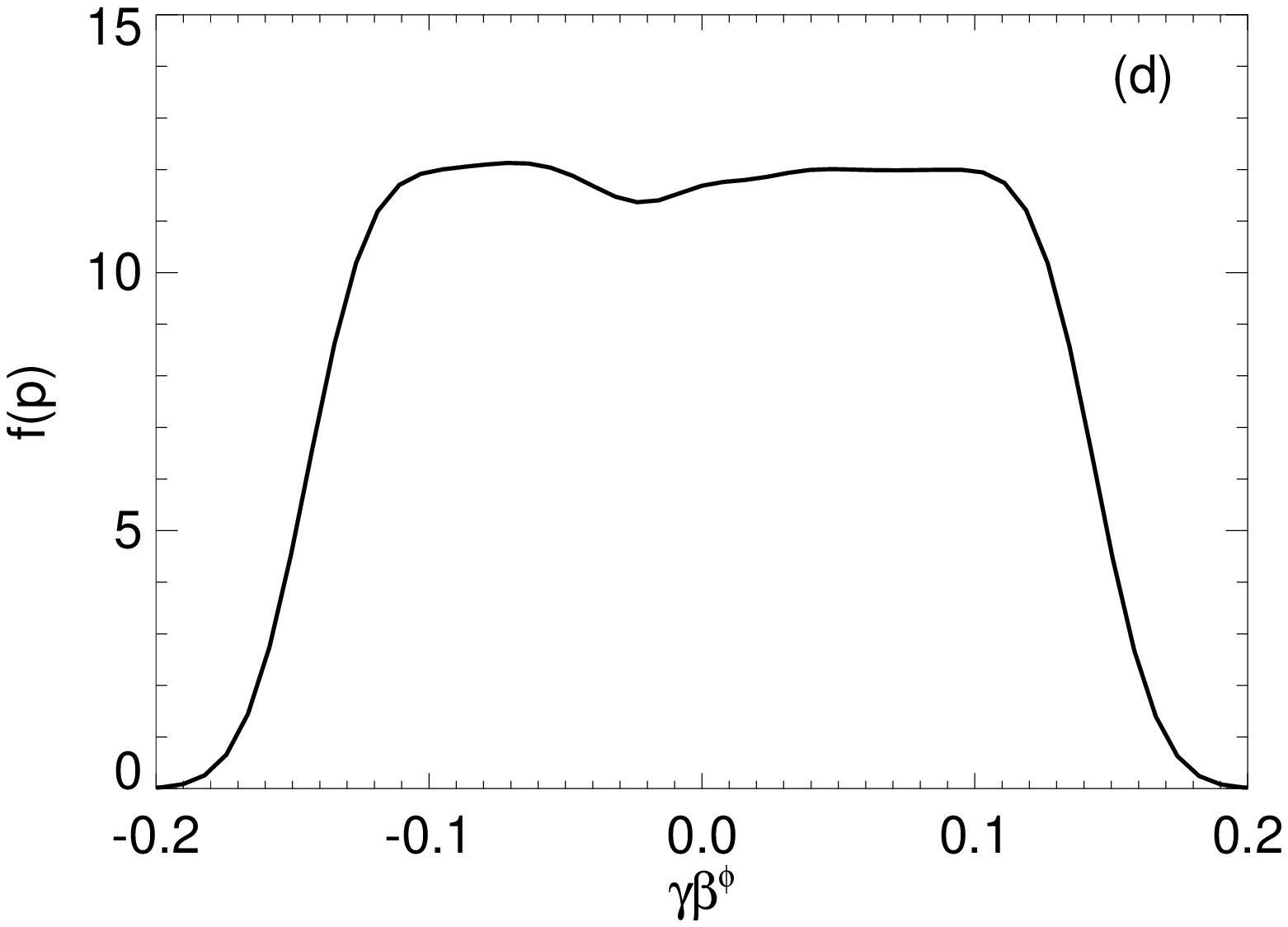}}
\end{center}
\end{figure*}

In Figure \ref{fig:df_100} we plot the momentum distributions of
unbound dark matter particles, as observed by a FFIO in the equatorial
plane, at a 
relatively large distance from the black hole: $r=100M$. Each
1-dimensional distribution is calculated by integrating over the other
two momentum dimensions. We also plot the momentum magnitude
$\gamma|\beta|$ in panel ({\it a}). Because the particles all have
relatively small velocities at infinity $\beta_0 \approx \sigma_0/c \ll
1$, their velocities in the weakly relativistic region $r_g \ll r \ll
r_0$ are given by $v\approx \sqrt{2GM/r}$, corresponding to $v\approx
0.14c$ for $r=100M$.

For the three spatial components of the momentum distribution, we see
a nearly isotropic velocity distribution with a few subtle but
interesting deviations. First, we note how there is a slight deficit
of particles with positive $p^{\tilde{r}}$. This is due to
capture by the black hole of particles coming in from infinity with
nearly radial trajectories. By definition, these particles also have
small values of $p^{\tilde{\theta}}$ and $p^{\tilde{\phi}}$, depleting
the distribution function in those dimensions around $\beta=0$. While
the distribution in the $\theta$ dimension is symmetric, note that the
depletion in the $\phi$ distribution is offset to slightly negative
values of $p^{\tilde{\phi}}$. This is due to the well-known
preferential capture by Kerr black holes of retrograde particles with
angular momenta aligned opposite to the black hole spin. 

\begin{figure*}[ht]
\caption{\label{fig:df_2} Momentum distribution of unbound
  particles observed by a FFIO in the equatorial plane at radius
  $r=2M$. Unlike Figure \ref{fig:df_100}, here we see a decidedly
  non-thermal and highly anisotropic distribution.}
\begin{center}
\scalebox{0.45}{\includegraphics*{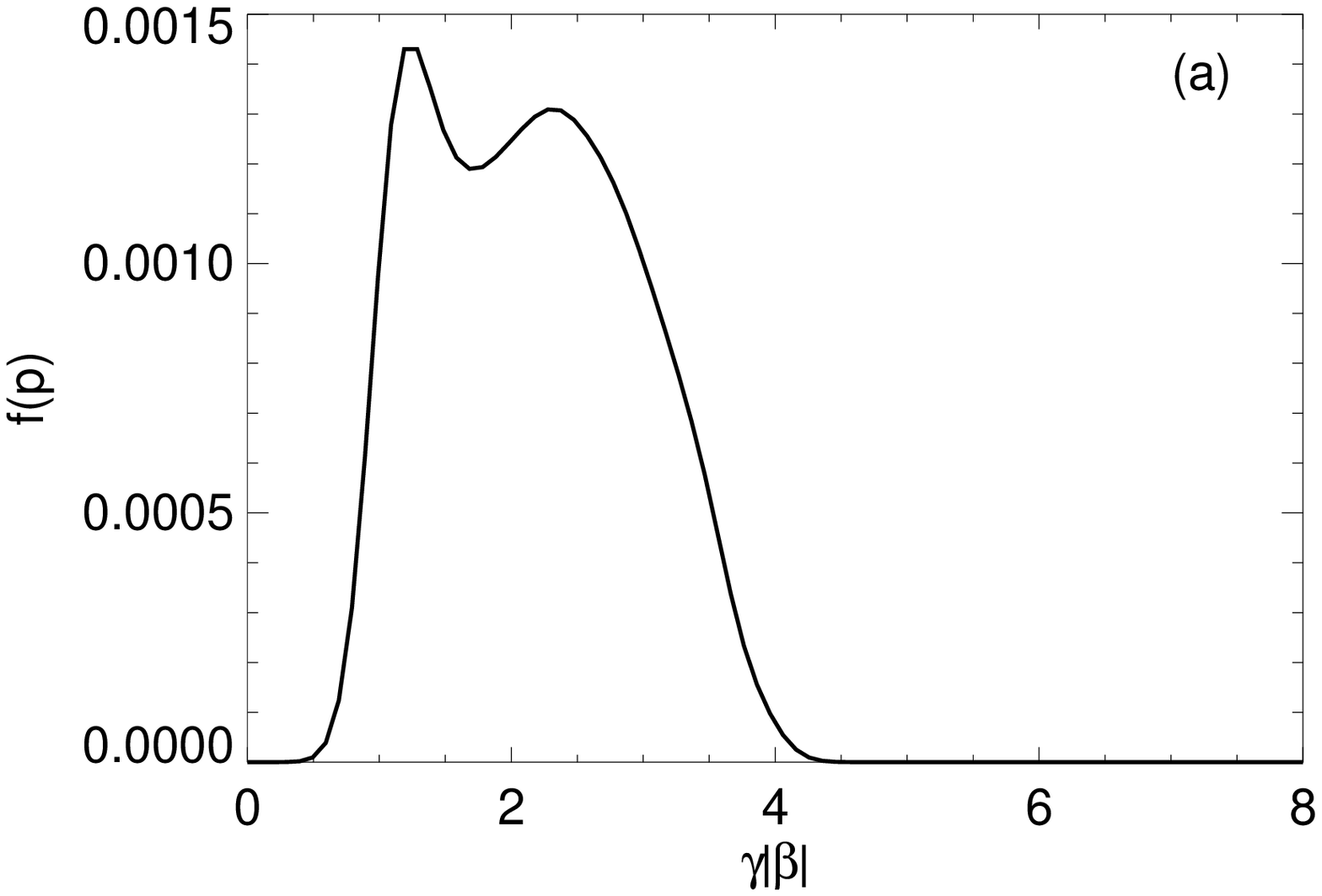}}
\scalebox{0.45}{\includegraphics*{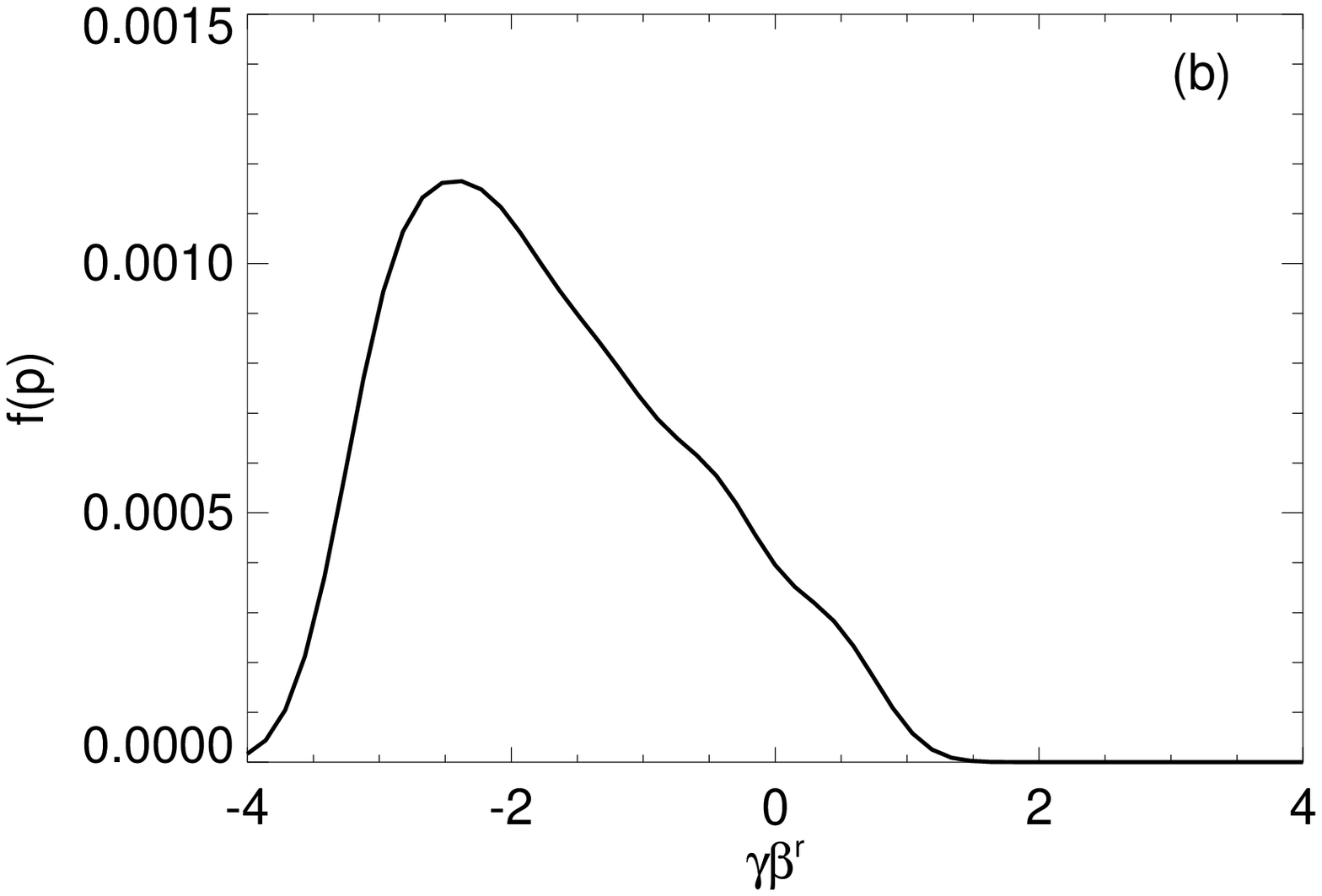}} \\
\scalebox{0.45}{\includegraphics*{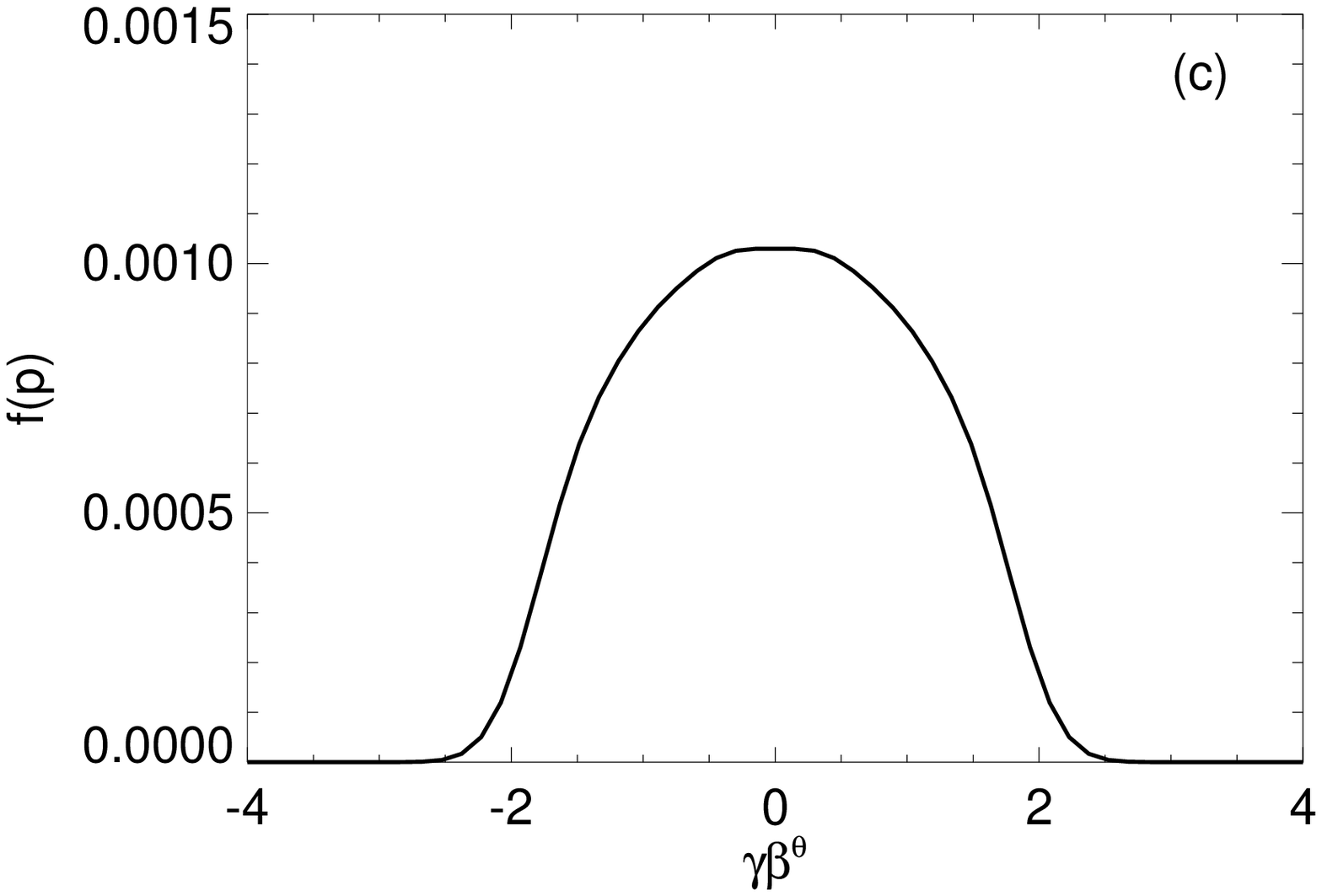}}
\scalebox{0.45}{\includegraphics*{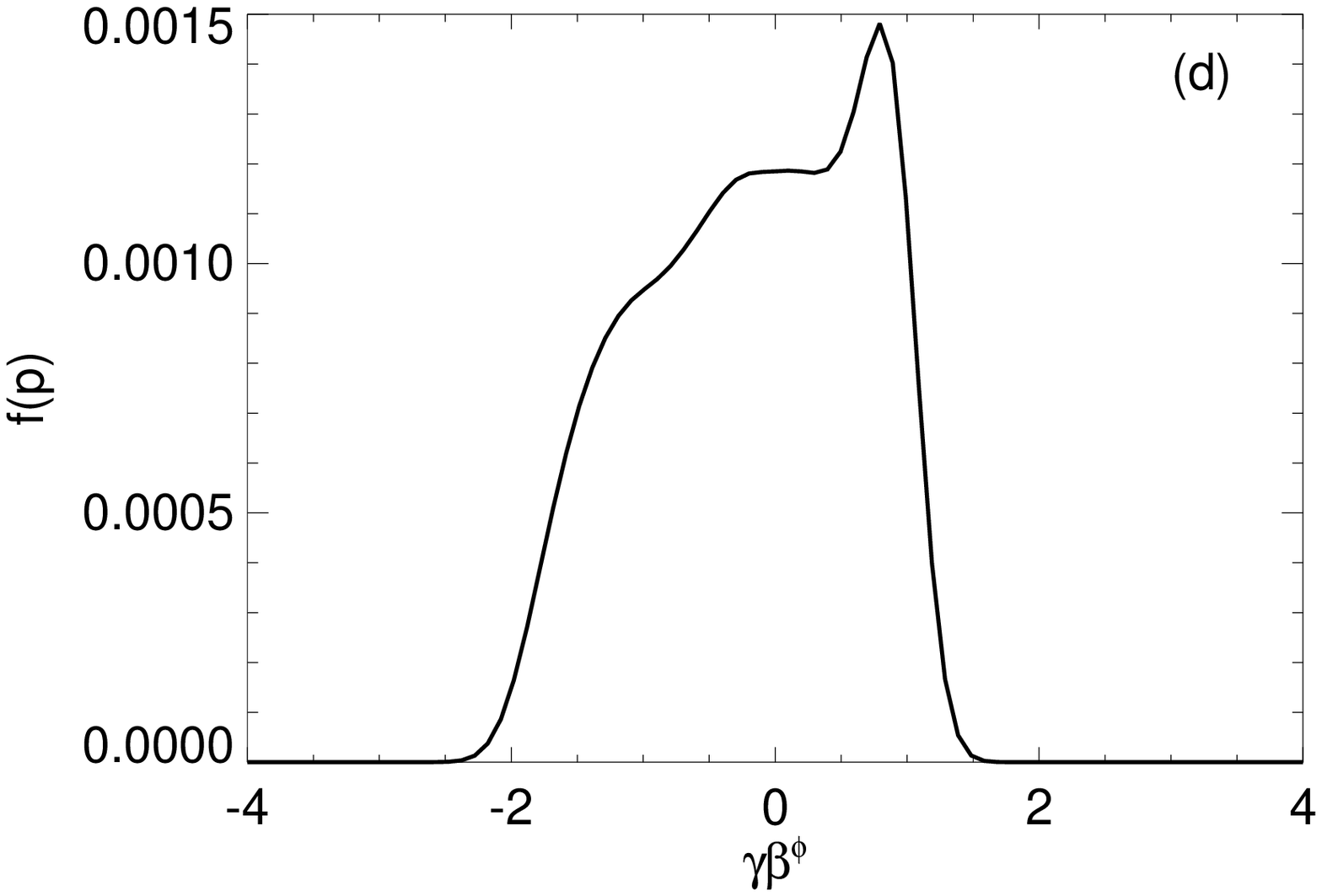}}
\end{center}
\end{figure*}

In Figure \ref{fig:df_2}, we plot the phase-space distribution for the
same boundary conditions as in Figure \ref{fig:df_100}, but now at
$r=2M$. The difference is quite dramatic, but all the features are
essentially due to the same physical mechanisms. This close to the
horizon, there is a very strong depletion of outgoing particles with
$p^{\tilde{r}}>0$, as most particles are captured by the black
hole. The only particles that can avoid capture at this radius have
prograde trajectories in the equatorial plane. Thus, the
distribution is now {\it peaked} around $p^{\tilde{\theta}}=0$ instead
of showing a deficit. 

There is also a strong peak near $p^{\tilde{\phi}}=1$ due to the
relatively stable, long-lived prograde orbits that circle the black hole
multiple times before getting captured or escaping back out to
infinity. In fact, the distribution of {\it coordinate} momentum
is significantly more lopsided to $p^\phi>0$, but this is masked in
Figure \ref{fig:df_2}d because this distribution is measured by an
observer with $u^\phi>0$ herself. The sharp fall-off of the
azimuthal distribution above $p^{\tilde{\phi}}\approx 1$ is due to the angular
momentum barrier of the black hole. Particles with higher values of
$p^{\tilde{\phi}}$ simply never reach this small radius. 

\begin{figure}[h]
\caption{\label{fig:Baushev} Comparison of our numerical results ({\it
    red}) with the analytic expression ({\it black}) for the particle
  density derived by \citet{Baushev2009} for a Schwarzschild black
  hole. The density here is defined in the coordinate, not proper,
  frame, leading to a much steeper rise at small $r$. In
  Boyer-Lindquist coordinates, the horizon for a non-spinning black
  hole is at $r=2M$.}
\begin{center}
\scalebox{0.45}{\includegraphics*{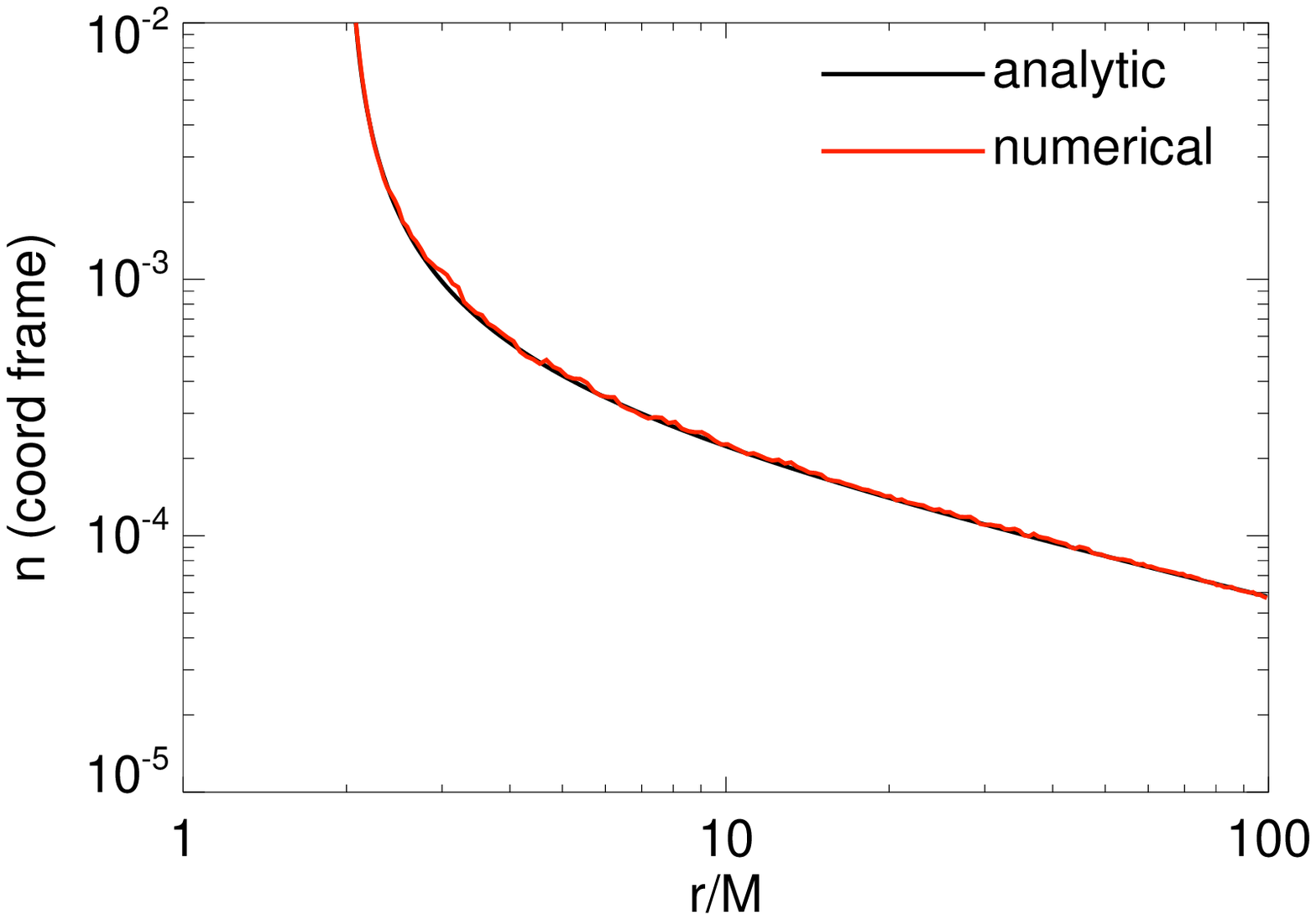}}
\end{center}
\end{figure}

To the best of our knowledge, these distribution functions have
never been calculated before for a Kerr black
hole. However, the particle number density can be determined analytically for a
non-spinning Schwarzschild black hole, in the limit of $\sigma_0\ll c$.
This allows at least one test of our numerical methods, although
admittedly not a very strong one, as most of the interesting
features are related to the far more complicated orbits around a
spinning black hole. We follow the approach of \citet{Baushev2009}, who integrates
the distribution function with fixed energy, carefully setting the
angular momentum integration bounds based on which orbits are captured
from a given radius. The results are shown in Figure
\ref{fig:Baushev}, with our numerical calculation plotted as a red
curve and the analytic result in black, showing perfect
agreement. Note that Baushev's expression is given for a {\it coordinate}
density rather than a proper density, which also explains the sharper
peak at small $r$. 

\subsection{Bound Particles}\label{section:bound}

As mentioned above, the unbound population can be thought of as a
lower limit on the total DM density. There will also likely
be a substantial population of particles that are gravitationally
bound to the black hole. As described in \citet{Gondolo1999}, the
origin of the bound population is the adiabatic growth of the
supermassive black hole on a timescale much longer than the typical
orbital time. This physical mechanism can be understood as follows: as
a marginally unbound DM particle passes within $r_{\rm infl}$, a small
amount of baryonic matter is accreted into this region, deepening the potential
well just enough to capture the particle onto a marginally {\it bound}
orbit. Once captured, the particle continues to orbit the black hole
while conserving its orbital angular momentum as the black hole
continues to gain mass. This has the effect of shrinking the radius of
the orbit.

Over time, more particles are captured and subsequently migrate closer
to the black hole, building up a steep density spike
\citep{Gondolo1999}. Inside of the inner-most stable circular orbit
(ISCO), there is a sharp falloff in the density spike due to plunging
trajectories \citep{Sadeghian2013}. Here we do not attempt to solve
for the slope of the density spike at large radii but leave it as a free 
parameter, and fix the density at the influence radius as for the
unbound population: $n_{\rm bound}(r) =  n_0
(r/r_0)^{-\alpha}$. Following \citet{Gondolo1999}, we also allow for 
the possibility of a density upper bound $n_{\rm annih}$
due to annihilation losses occurring over very long timescales. 

To populate the phase-space distribution for the bound population,
we follow a similar method as described above for the unbound
particles, but instead of launching them from large radius with a
limited range of impact parameters, now we launch them {\it in situ}
with a isotropic thermal velocity distribution, as measured by a local
ZAMO. These particles begin much closer to the black hole, so the
relativistic Maxwell-J{\"u}ttner velocity distribution is used
\citep{Juttner1911}, with the characteristic virial temperature
$\Theta(r)=1/2[1-\epsilon_{\rm ZAMO}(r)]$, where $\epsilon_{\rm
  ZAMO}(r)-1$ is the specific gravitational binding energy of the
ZAMO.

Because many of the particles launched close to the black hole get
captured, we first integrate their trajectories for a few orbital
periods to ensure they are in fact on stable orbits. Only then
do they contribute to the tabulated distribution function. 
Additionally, a small fraction of the test particles from the tail end
of the velocity distribution will in fact be unbound, and these are
similarly discarded. 

As with the unbound distribution, for each step
along its trajectory, the test particle
contributes to the phase space distribution a small weight
proportional to the amount of time spent on that step. Yet now,
instead of using the coordinate time $dt$, we use the proper time of
the ZAMO frame from which the particles are launched, including an
additional weight to ensure the appropriate radial form of the
density distribution at larger radii.

\begin{figure}[h]
\caption{\label{fig:dn_dr_bound} Spatial density ({\it a}) and mean
  relative momentum $\langle\gamma\beta\rangle_{\rm rel}$ ({\it b}) of
  bound particles in the equatorial plane of a Kerr black hole with
  $a/M=1$, as measured in the ZAMO frame. The dashed lines are the
  Newtonian solutions when $n(r)\sim r^{-2}$ far from the black
  hole.}
\begin{center}
\scalebox{0.45}{\includegraphics*{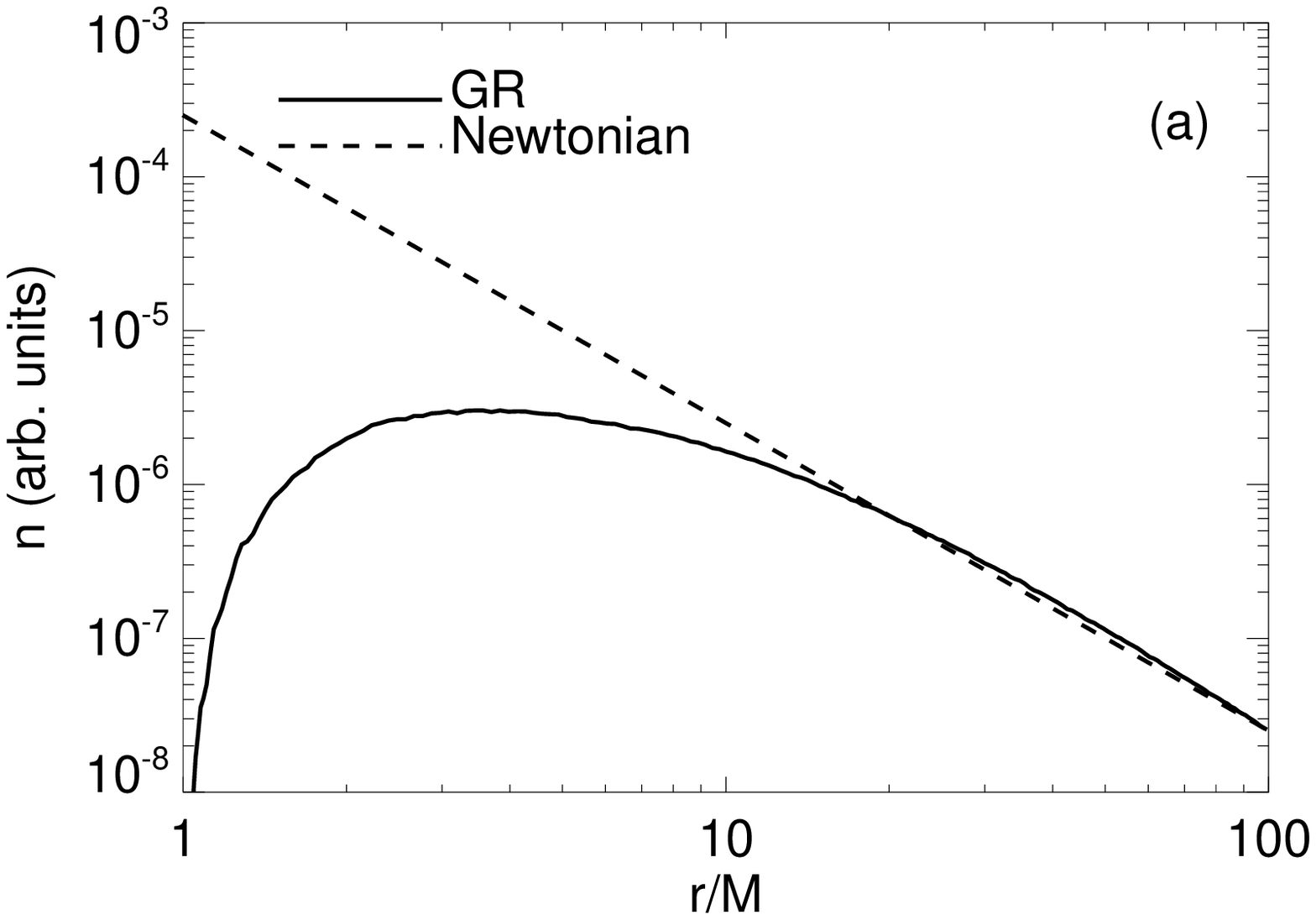}}
\scalebox{0.45}{\includegraphics*{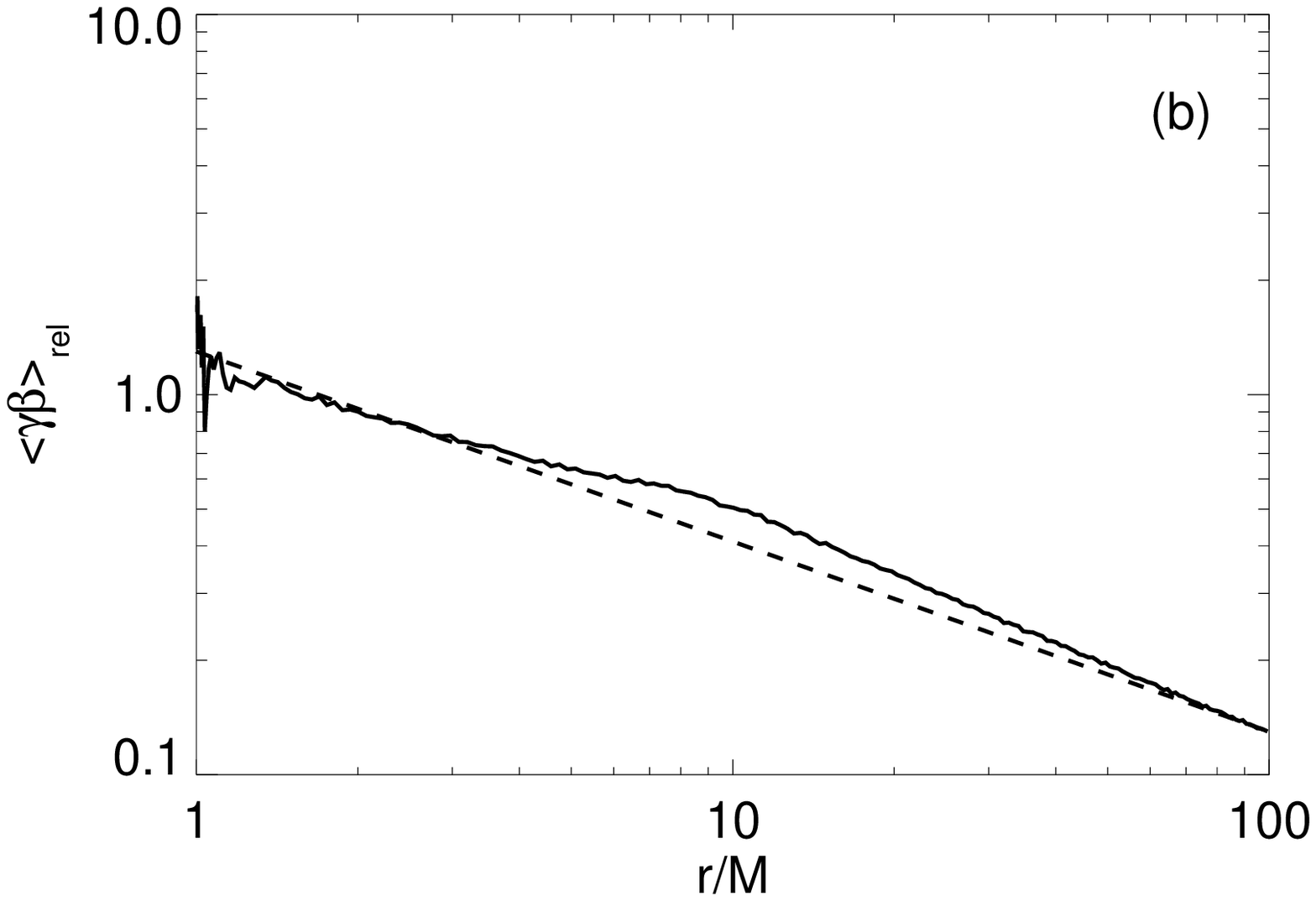}}
\end{center}
\end{figure}

In Figure \ref{fig:dn_dr_bound} we plot the radial density
distribution and mean relative momentum of the bound particles, as
measured in the ZAMO frame, in the equatorial plane around a Kerr
black hole with spin $a/M=1$.
The density profile is constructed so that
$\rho(r)\sim r^{-2}$ at large radii. We clearly see major differences
relative to the unbound population shown in Figure
\ref{fig:dn_dr}. Because of the lack of stable orbits close to the
black hole, the bound population declines inside $r\approx 4M$, which
corresponds roughly to the mean radius of the ISCO for randomly
inclined orbits around a maximally spinning black hole. This effect
was described in detail for
non-spinning black holes in \citet{Sadeghian2013}. For equatorial
circular orbits, only prograde trajectories are allowed inside of
$r=9M$. This leads to all particles moving in roughly the same
direction closer to the black hole, and explains why the relative
momentum $\langle \gamma\beta\rangle_{\rm rel}$ does not increase
nearly as fast for the bound population as it does for the unbound
population, which allows plunging retrograde trajectories, and thus
more ``head-on'' collisions. 

\begin{figure*}[ht]
\caption{\label{fig:n_rth} Spatial density of test particles in
  the $x-z$ plane, for both bound and unbound populations, for $a/M=0$
  and $a/M=1$. For each case, we show the
  unbound distribution on the left side and the bound distribution on
  the right side of the plot, and all distribution functions are
  normalized to the mean density at $r=10M$.
  The horizon is plotted as a solid curve
  and the radius of the marginally bound orbits is shown as a dotted
  curve. The spin axis of the black hole is in the $+z$ direction.}
\begin{center}
\scalebox{0.45}{\includegraphics*{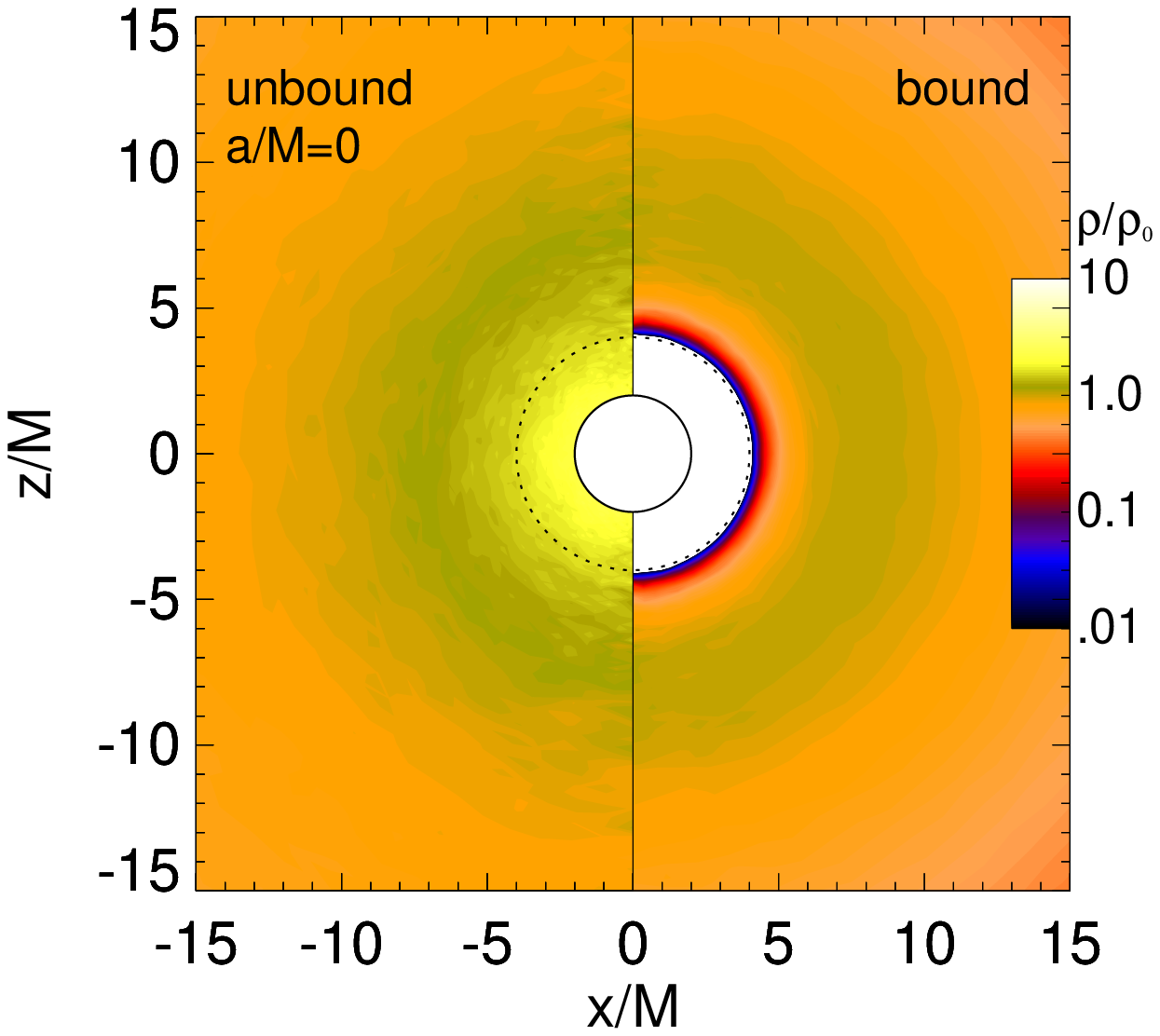}}
\scalebox{0.45}{\includegraphics*{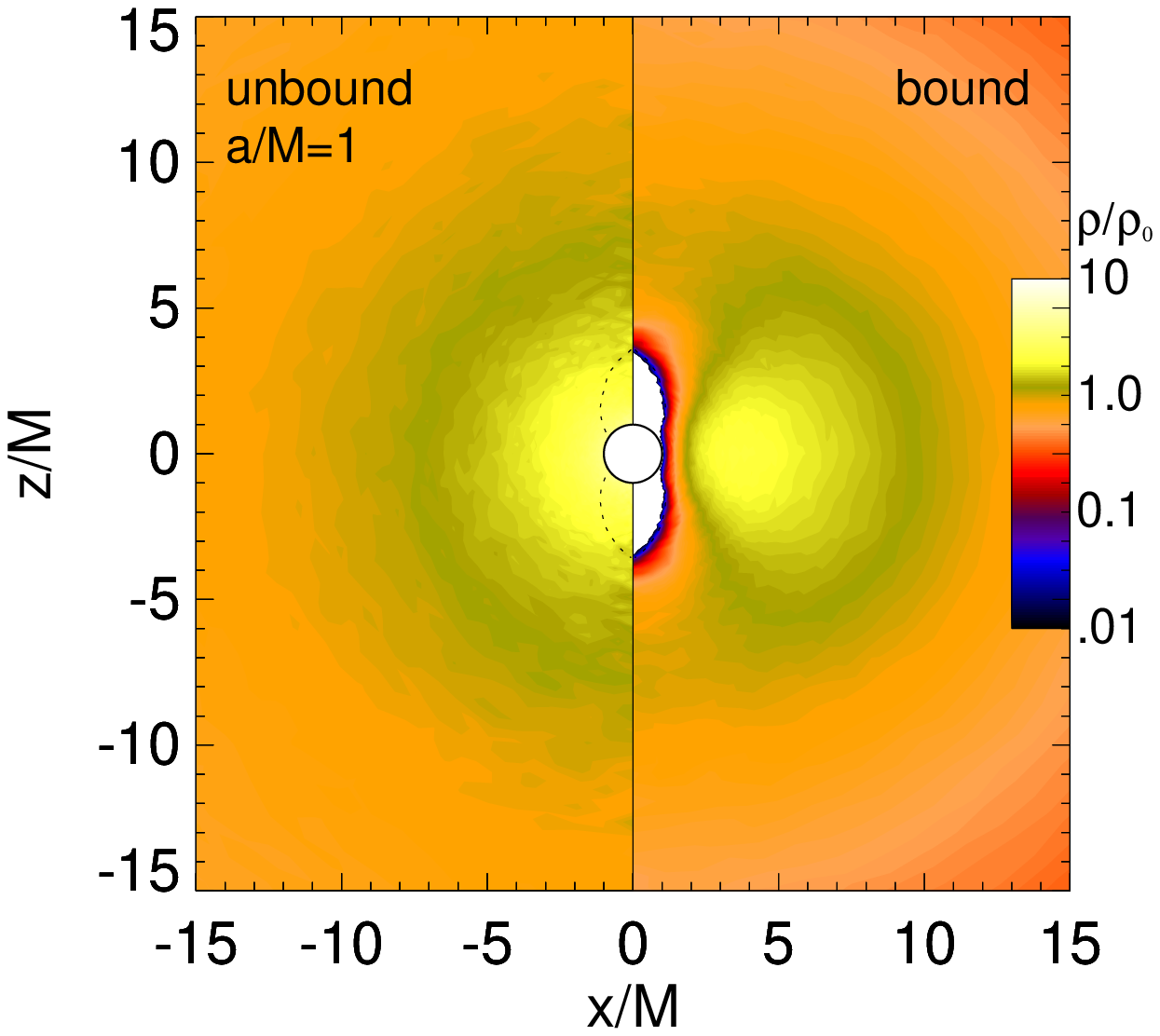}}
\end{center}
\end{figure*}

In Figure \ref{fig:n_rth} we show the 2D density profile in the $x-z$
plane for both bound and unbound populations, for $a/M=0$ and
$a/M=1$. The horizon in Boyer-Lindquist coordinates is
plotted as a solid black line. 
For comparison purposes, the density scale is normalized to the
mean value at $r=10M$. In reality, the density of the bound particles
could be orders of magnitude greater at these radii
\citep{Gondolo1999}. The most obvious difference here is
the depletion of bound orbits inside of the ISCO, which lies at $r=6M$
for non-spinning black holes. For spinning black
holes, the radius of the ISCO is a function of the particle's
inclination angle, ranging from $r=1M$ for prograde orbits in the
equatorial plane, to $r= 5.2M$ for polar orbits, and $r=9M$ for
retrograde equatorial orbits. 

Inside of the ISCO, there is also the ``marginally bound'' radius,
where particles with unity specific energy can exist on unstable
circular orbits. This radius is also a function of inclination angle,
and is plotted in Figure \ref{fig:n_rth} as dotted curve. Inside of
this orbit, no bound particles will be found (for improved visibility,
we have left this region white, not black, as would be required by a
strict adherence to the color scale). One interesting feature
of Figure \ref{fig:n_rth} is that the density of the {\it unbound}
population around spinning black holes doesn't show any obvious
$\theta$-dependence. It appears that the enhanced density due to
long-lived prograde orbits is almost exactly countered by the lack of
retrograde orbits at the same latitude. 

\begin{figure*}[ht]
\caption{\label{fig:df_100_bound} Momentum distribution of bound
  particles observed by a ZAMO in the equatorial plane at radius
  $r=100M$. Unlike the unbound distribution in Figure \ref{fig:df_100},
  the energy distribution is much broader here, yet with a smaller mean
  momentum (panel {\it a}). In panels ({\it
    b-d}) we show the distribution of the individual momentum
  components.} 
\begin{center}
\scalebox{0.45}{\includegraphics*{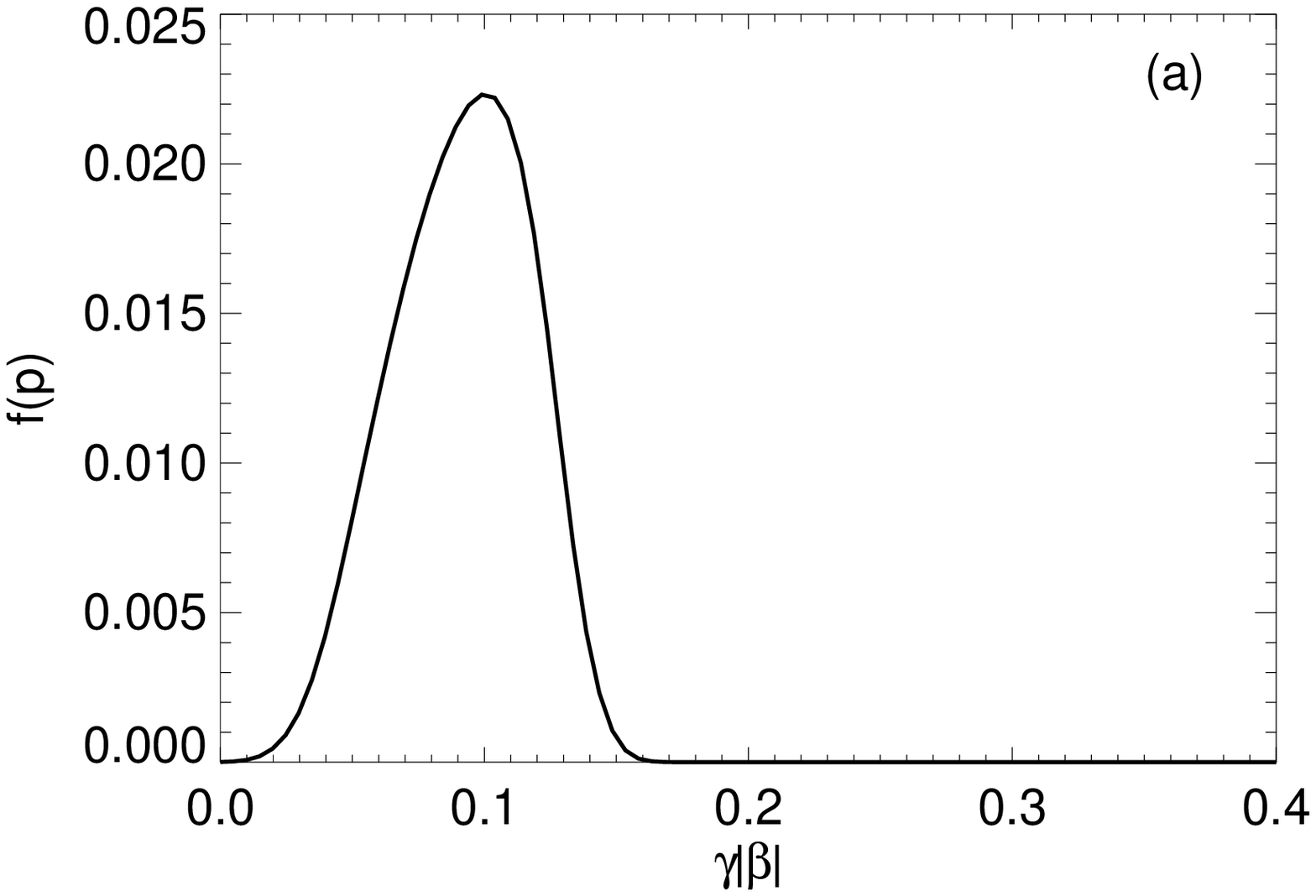}}
\scalebox{0.45}{\includegraphics*{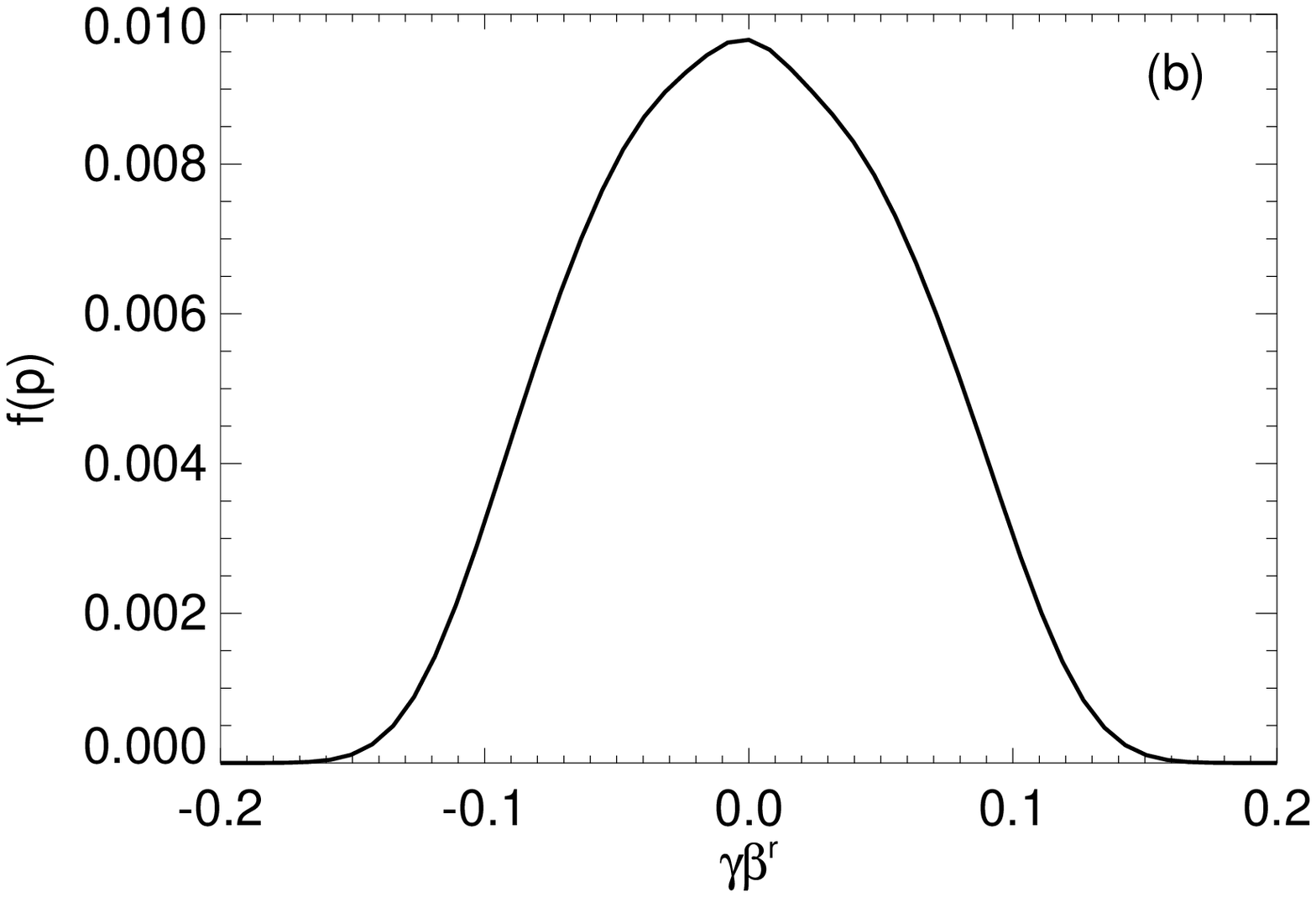}} \\
\scalebox{0.45}{\includegraphics*{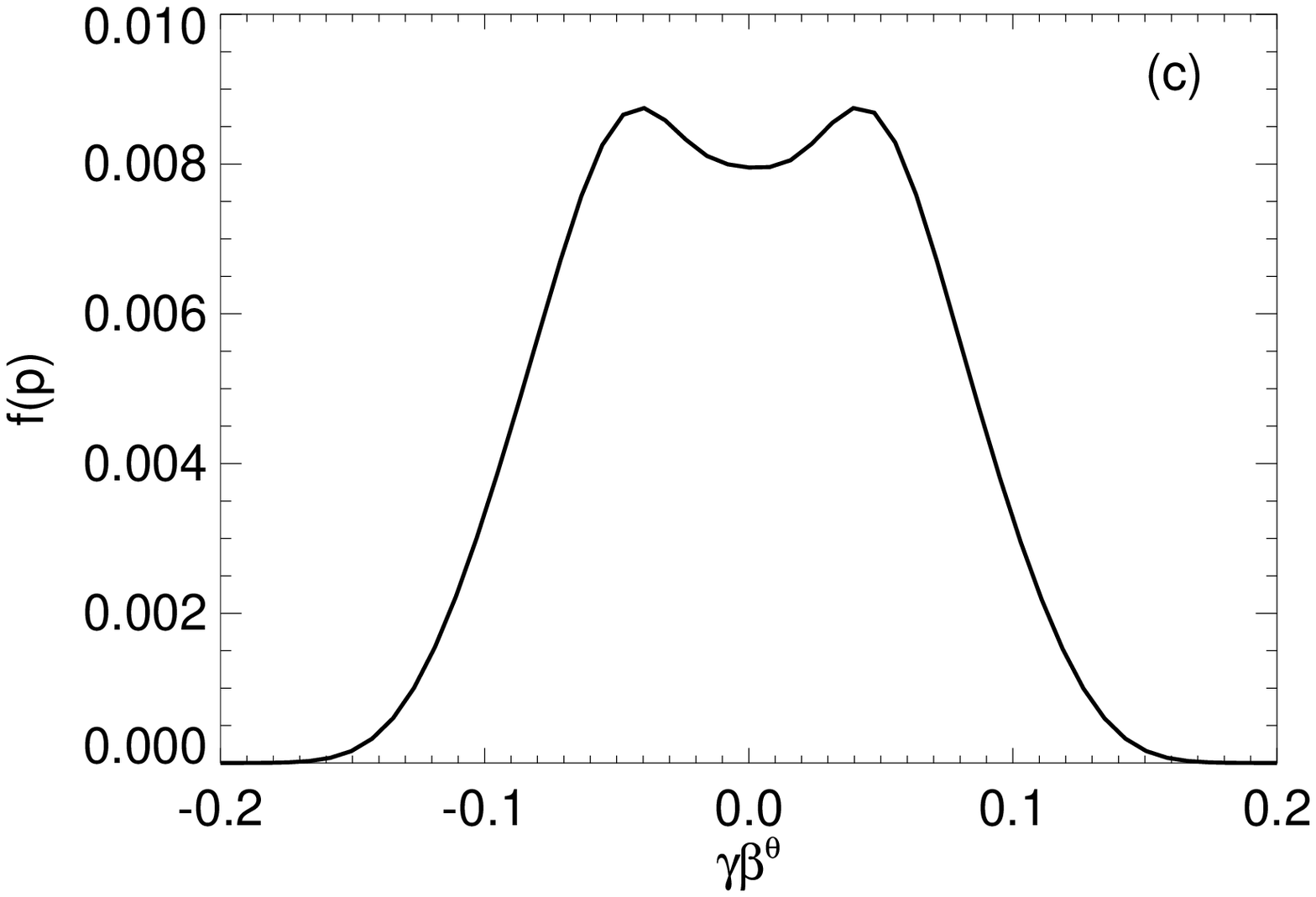}}
\scalebox{0.45}{\includegraphics*{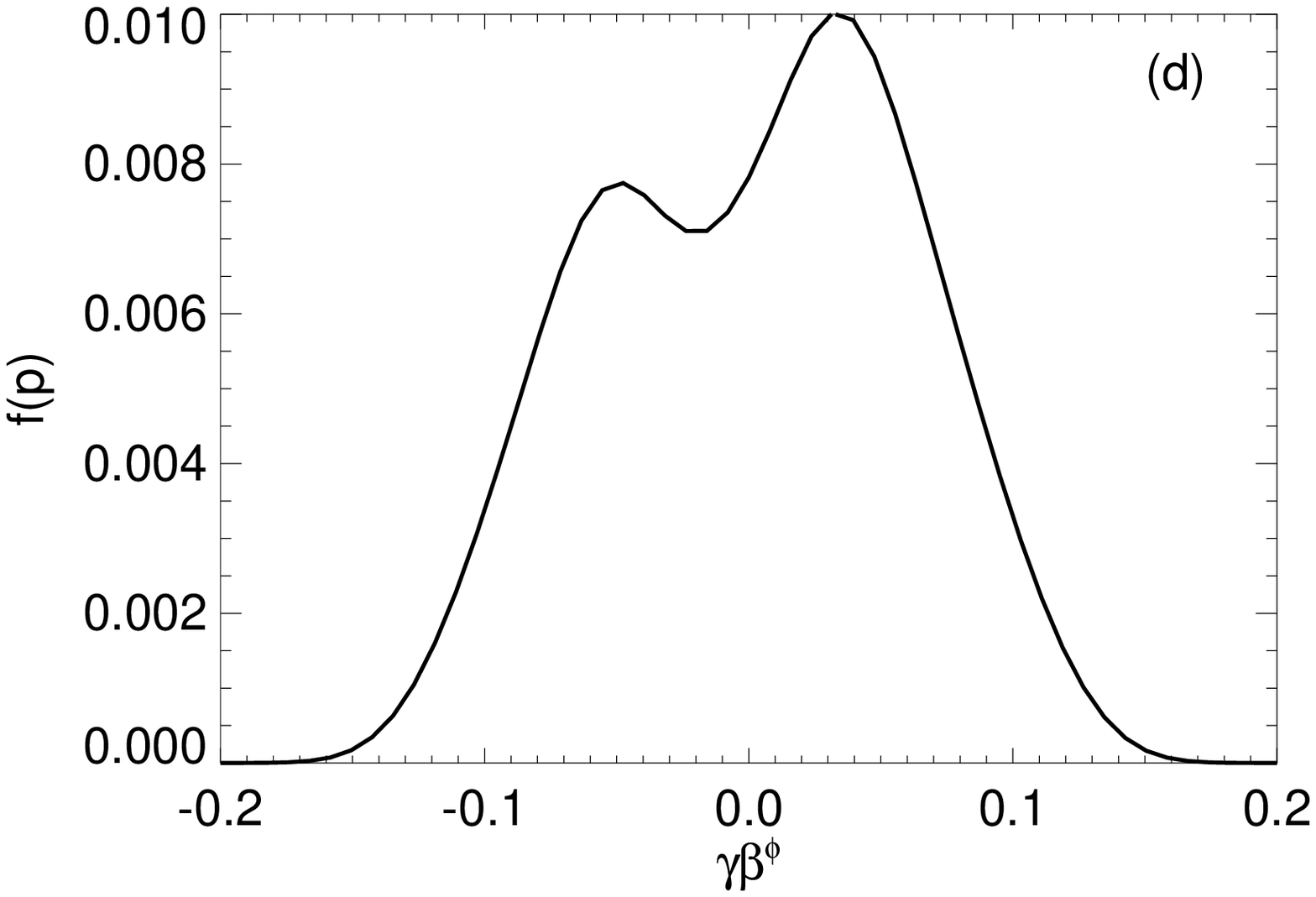}}
\end{center}
\end{figure*}

In Figure \ref{fig:df_100_bound} we show the phase space distribution
for each of the momentum components, as measured by a ZAMO in
the equatorial plane at large radius ($r=100M$). Compared to the
equivalent plot for the unbound distribution (Fig.\ \ref{fig:df_100}),
we see a number of significant differences. First, the 
fact that these particles are bound requires that $E<1$, and the
imposed virial energy distribution results in mean velocities that are
smaller than those of the unbound population by a factor of
$\sim\sqrt{2}$. Second, because we require stable, long-lived orbits,
there is a larger depletion around $p^{\tilde{\theta}}=0$ and $p^{\tilde{\phi}}=0$,
as these trajectories are all captured by the black hole and thus do
not contribute at all to the distribution function. Similarly,
we see a larger asymmetry due to the preferential capture of
retrograde orbits with $p^{\tilde{\phi}}<0$. 

\begin{figure*}[ht]
\caption{\label{fig:df_2_bound} Momentum distribution of bound
  particles observed by a ZAMO in the equatorial plane at radius
  $r=2M$. Compared to Figure \ref{fig:df_2}, here we actually
  see a {\it more}
  symmetric, thermal distribution making up a thick torus of stable,
  roughly circular orbits near the equatorial plane.}
\begin{center}
\scalebox{0.45}{\includegraphics*{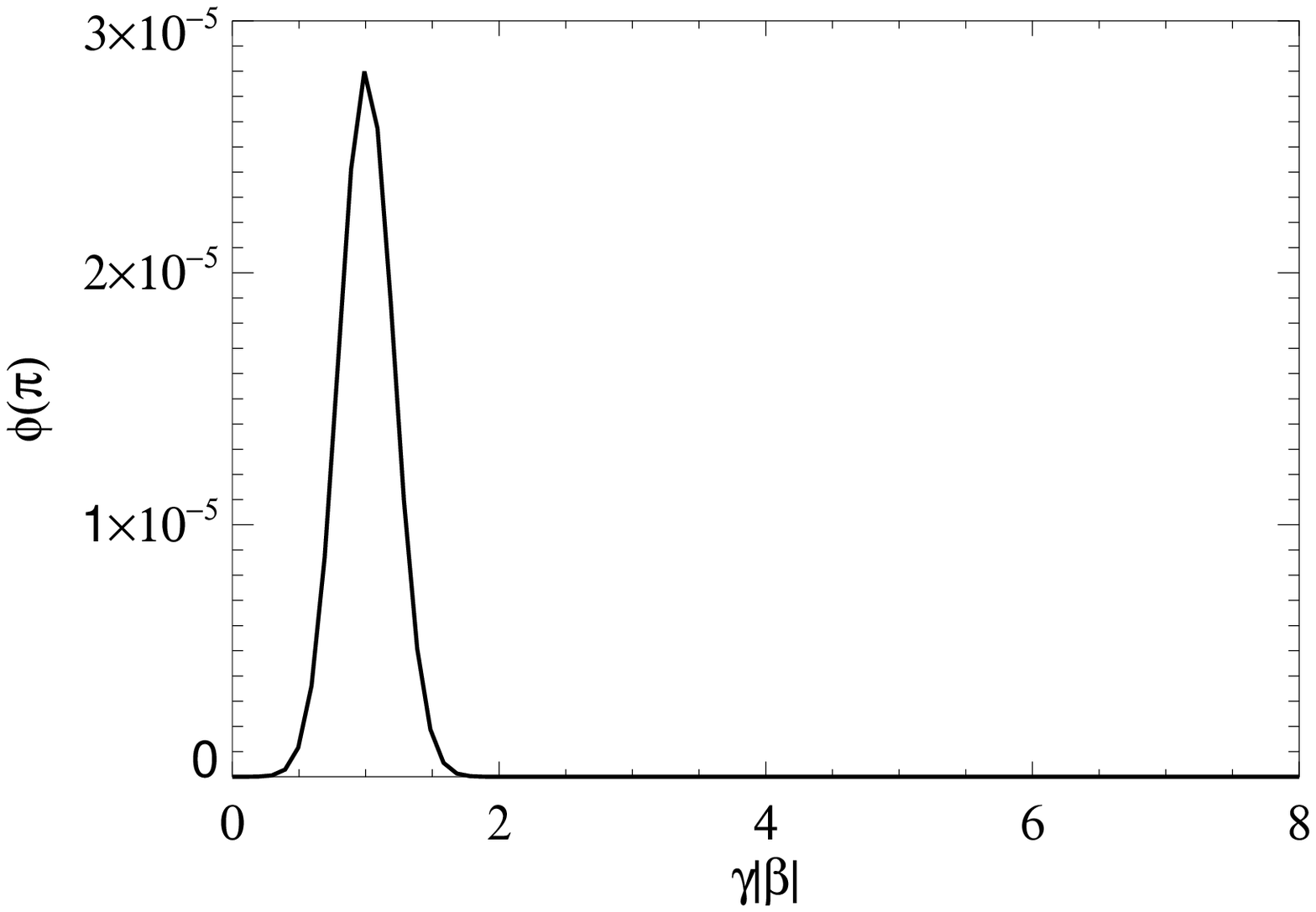}}
\scalebox{0.45}{\includegraphics*{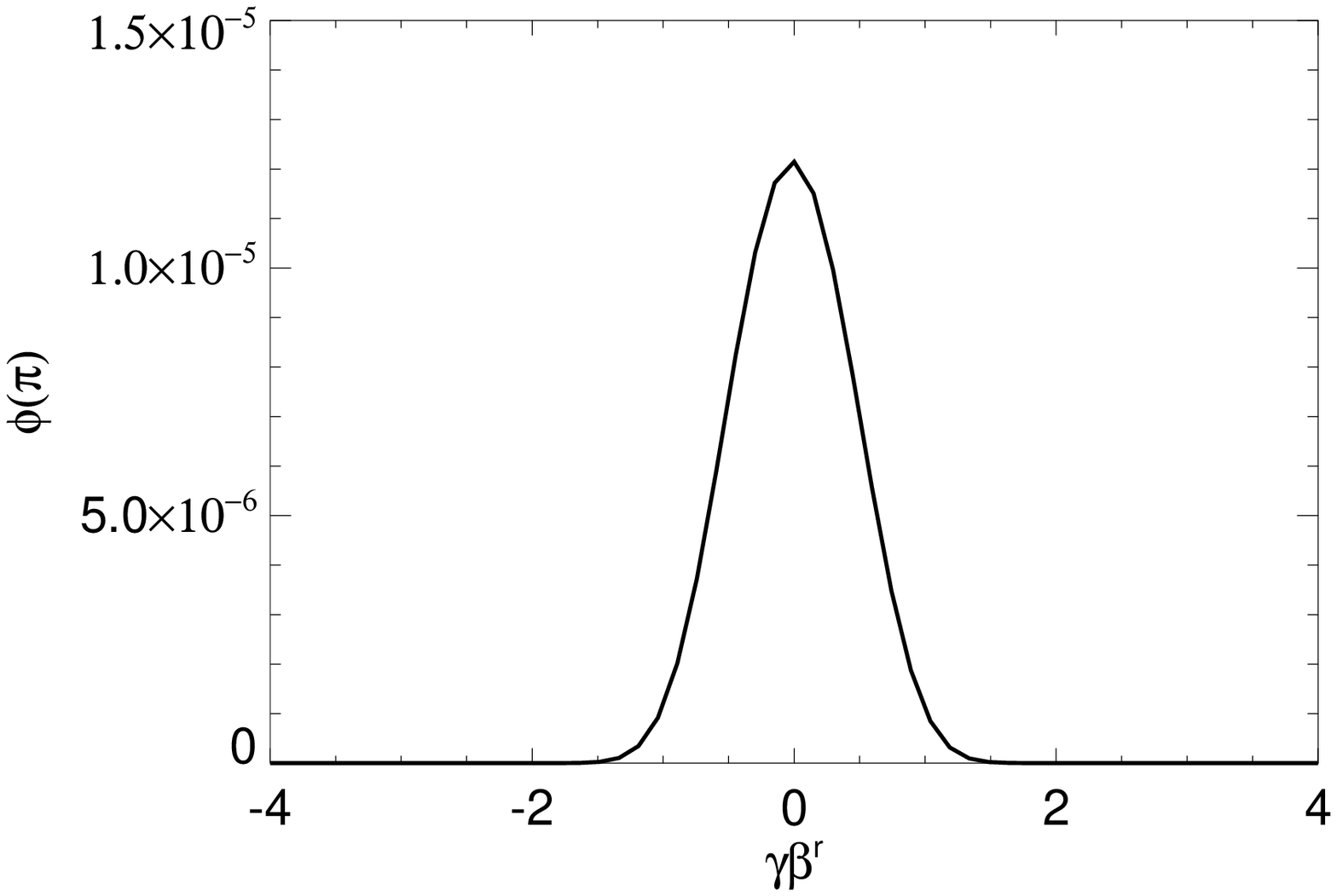}} \\
\scalebox{0.45}{\includegraphics*{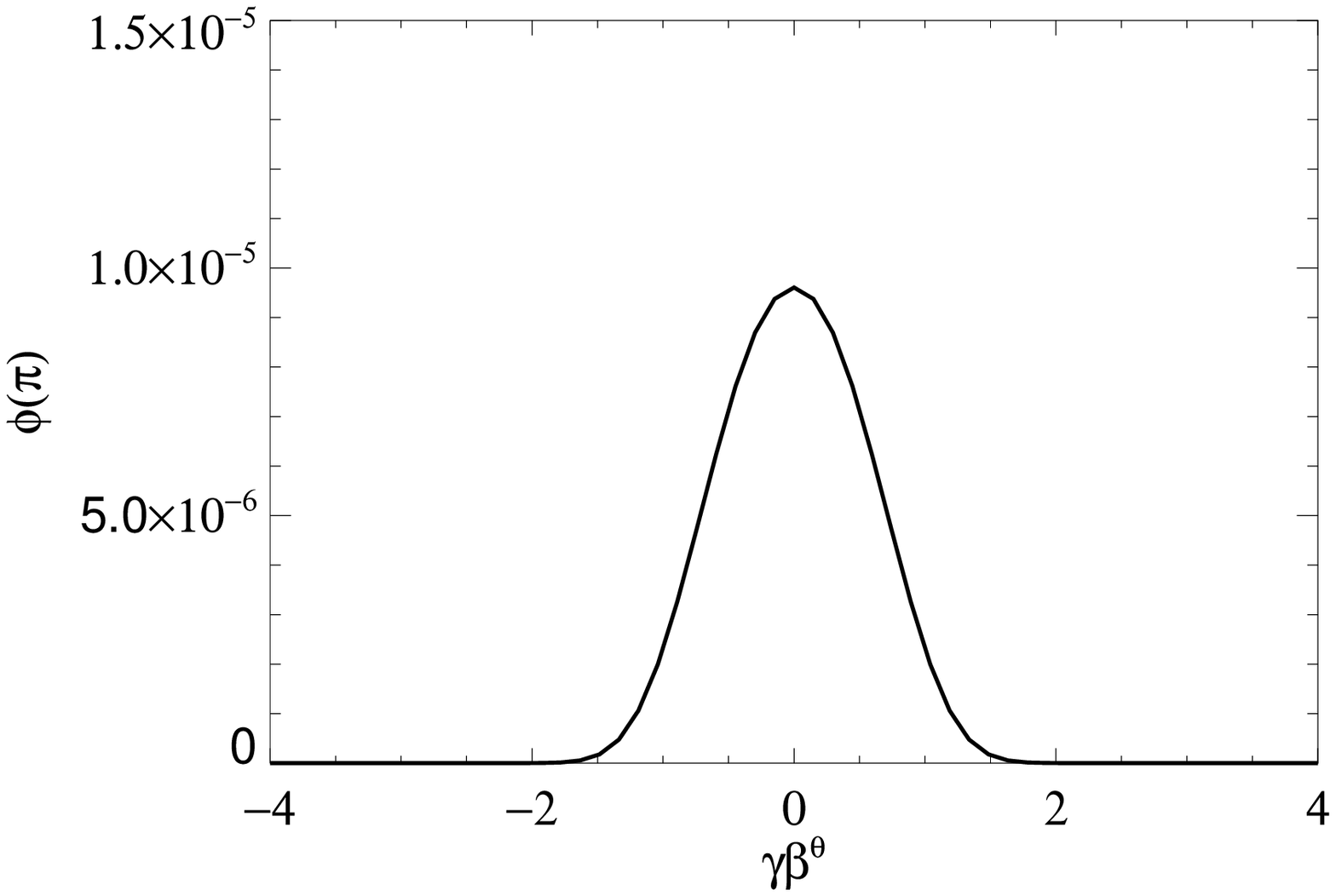}}
\scalebox{0.45}{\includegraphics*{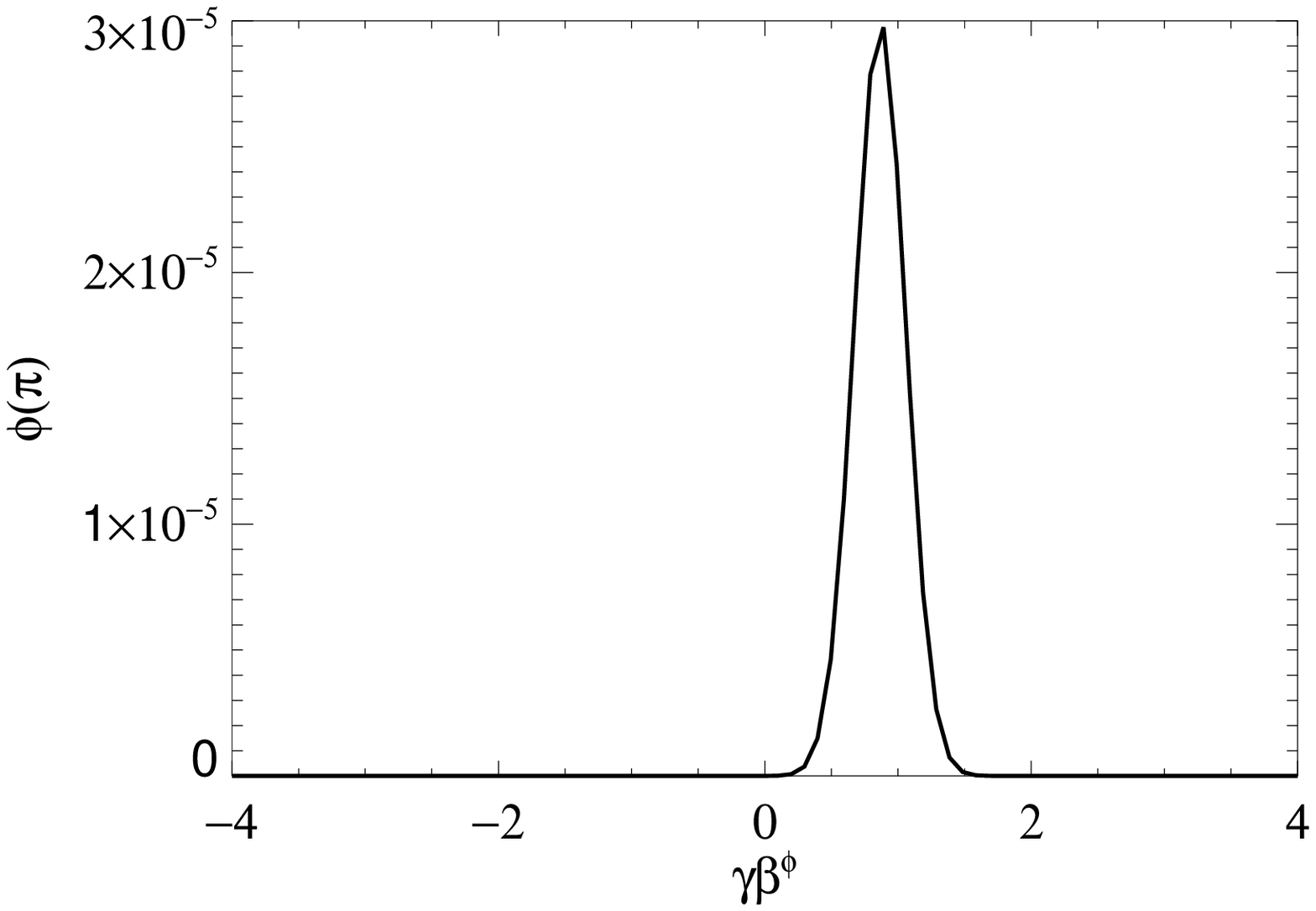}}
\end{center}
\end{figure*}

In Figure \ref{fig:df_2_bound} we plot the same momentum distribution
functions, now at $r=2M$. Here the contrast with the unbound
population (Fig.\ \ref{fig:df_2}) is even greater. The only stable
orbits at this radius are prograde, nearly circular, nearly equatorial
orbits. This results in a relatively narrow distribution
clustered around $u^{\tilde{\mu}} = [\sqrt{2},0,0,1]$ in the
ZAMO frame. This narrower range in allowed velocities will have a
profound impact on the shape of the annihilation spectrum, as we will
see in the following section.

\section{ANNIHILATION PRODUCTS}\label{section:raytracing} 

Once we have populated the distribution function, we can calculate the
annihilation rate given a simple particle-physics model for the dark
matter cross section. Again, it is simplest to work in the local
tetrad frame. Including special relativistic corrections
\citep{Weaver1976}, the local reaction rate is given by the following:
\begin{equation}\label{eqn:rate}
R(\mathbf{x}) = \int d^3 \mathbf{p}_1 \int d^3
\mathbf{p}_2\, f(\mathbf{x},\mathbf{p}_1)\, 
f(\mathbf{x},\mathbf{p}_2)\, 
\frac{\gamma_{\rm rel}}{\gamma_1 \gamma_2}\, \sigma_\chi(\gamma_{\rm
      rel})\, v_{\rm rel} \, ,
\end{equation}
where $\gamma_1$ and $\gamma_2$ are the Lorentz factors of two
particles as measured in the tetrad frame, $v_{\rm rel}$ is their
relative velocity, and $\sigma_\chi$ is the annihilation cross section
(potentially a function of the relative velocity). 
$R(\mathbf{x})$ has units of [events per unit proper volume per
  unit proper time], so we multiply by $d\tau/dt$ to get the rate
observed by a distant observer.

\begin{figure}[h]
\caption{\label{fig:reaction_schem} For a given phase-space
  distribution $f(\mathbf{x},\mathbf{p})$, the annihilation rate is
  calculated in each discrete volume element around the black
  hole. Every annihilation event samples the distribution function to
  get the momenta for the two dark matter particles
  $\mathbf{p}_1$ and $\mathbf{p}_2$ and produces two photons
  $\mathbf{k}_3$ and $\mathbf{k}_4$ with isotropic distribution
  in the center-of-mass frame. The product photons then propagate
  along geodesics until they reach a distant observer or get captured
  by the black hole.}
\begin{center}
\scalebox{0.45}{\includegraphics*{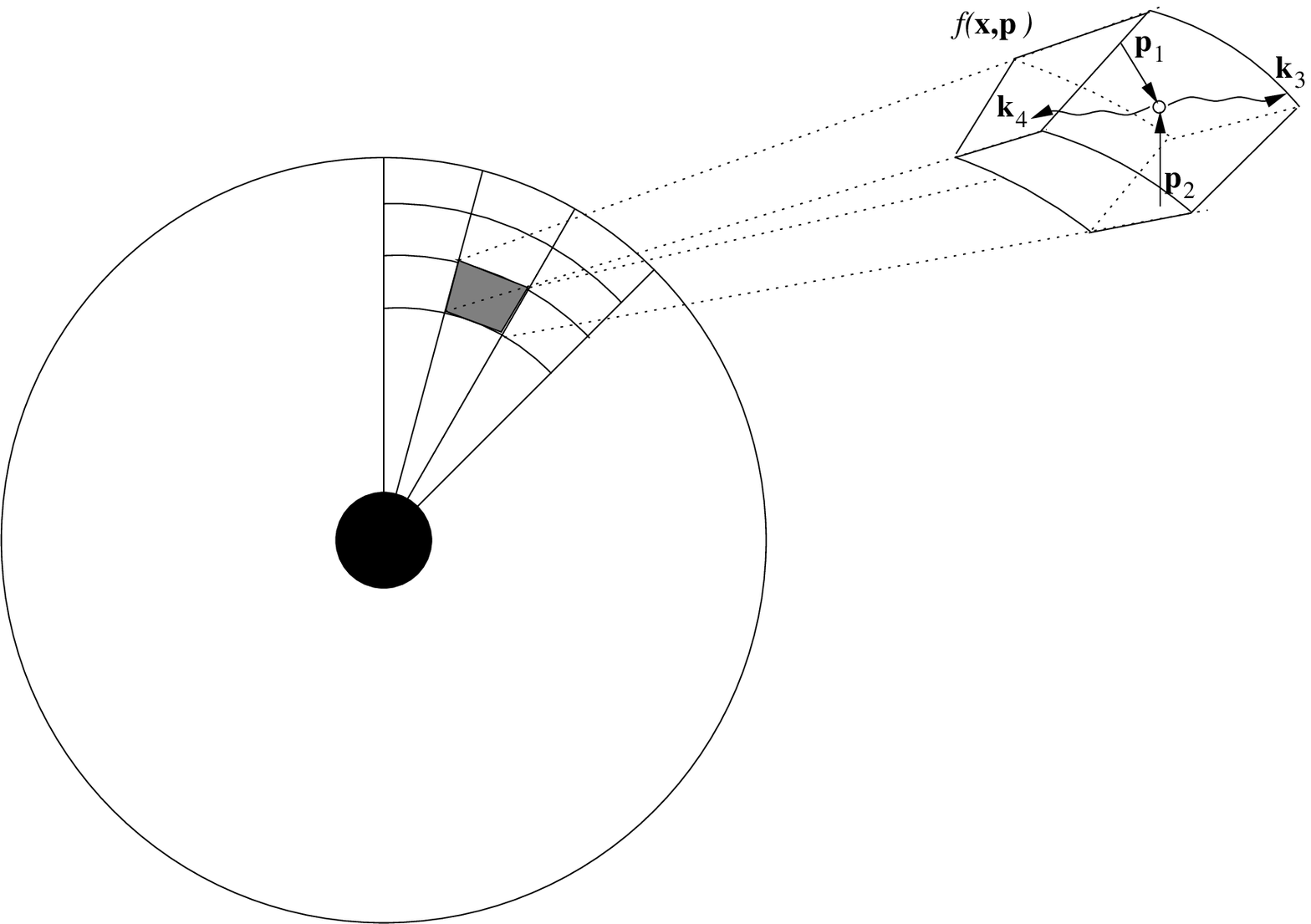}}
\end{center}
\end{figure}

The distribution function $f(\mathbf{x},\mathbf{p})$ is calculated
numerically using the methods of Section
\ref{section:df}. As discussed there, the numerical representation of
$f$ can have upwards of $10^8$ elements, so the direct integration of
equation (\ref{eqn:rate}) is generally not computationally feasible. Instead, we
use a Monte Carlo sampling algorithm to pick random momenta for each
particle with an appropriate weight based on the magnitude of $f$ and
the size of the discrete phase space volume. 

The spatial integration, however, is carried out directly, looping
over coordinates $r$ and $\theta$. This is shown schematically in
Figure \ref{fig:reaction_schem}. For each volume element, a large
number (typically $\sim 10^6$) of pairs of particles are sampled, and
for each pair, a center-of-mass tetrad is created. The total
energy in the center-of-mass frame is given by
\begin{equation}\label{eqn:E_com}
E_{\rm com}=m_\chi\sqrt{2(1+\mathbf{u}_1\cdot\mathbf{u}_2)}\, ,
\end{equation}
where $m_\chi$ is the rest mass of the DM particle, and
$\mathbf{u}=\mathbf{p}/m_\chi$ is the particle 4-velocity. 

The 4-velocity of the center-of-mass frame is then given by
\begin{equation}\label{eqn:u_com}
\mathbf{u}_{\rm com}=(\mathbf{u}_1+\mathbf{u}_2)/E_{\rm com}\, .
\end{equation}
The center-of-mass tetrad is constructed with 
$\mathbf{e}_{(\tilde{t})}=\mathbf{u}_{\rm com}$. The spatial
basis vectors are totally arbitrary, as they are only needed to
launch photons with an isotropic distribution in the center-of-mass
frame. Two photons, labeled $\mathbf{k}_3$ and $\mathbf{k}_4$ in
Figure \ref{fig:reaction_schem}, are launched in opposite directions,
each with energy $E_{\rm com}/2$ in the center-of-mass frame. We then
transform back to a coordinate basis for the geodesic integration of
the photon trajectories to a distant observer. 

\begin{figure}[h]
\caption{\label{fig:image_100} Simulated image and spectrum of the annihilation
  signal from unbound dark matter out to a radius $r=100M$ around a
  Kerr black hole. The
  observer is located in the equatorial plane. While the brightness
  peaks towards the black hole, the total flux is dominated by
  annihilations at large radii. The central shadow is
  clearly seen, blocking emission coming from the far side of the
  black hole. The photon energy $E$ is scaled to the
  dark matter rest mass $m_\chi$.}
\begin{center}
\scalebox{0.6}{\includegraphics*[55,450][360,720]{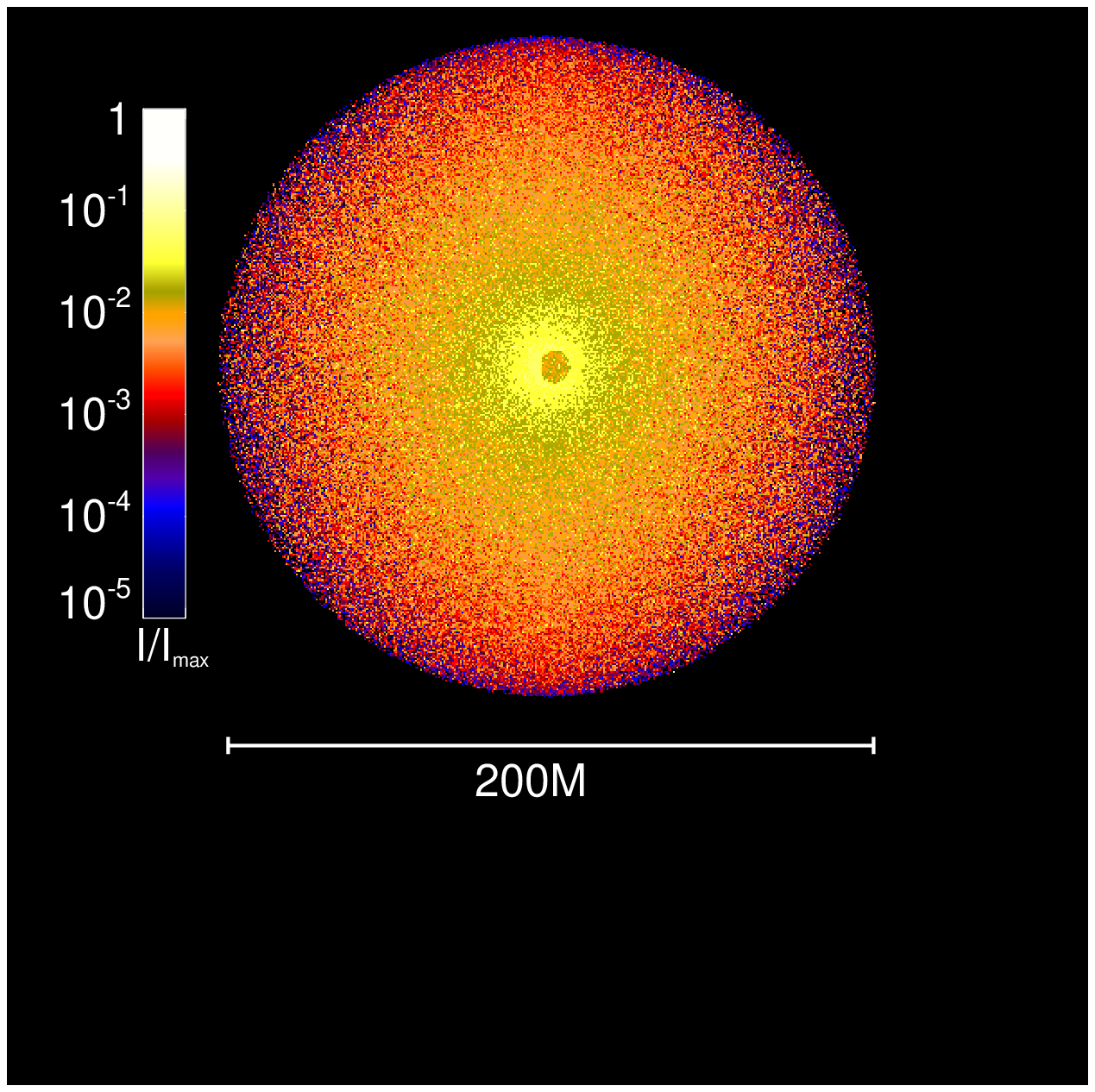}}
\scalebox{0.45}{\includegraphics*{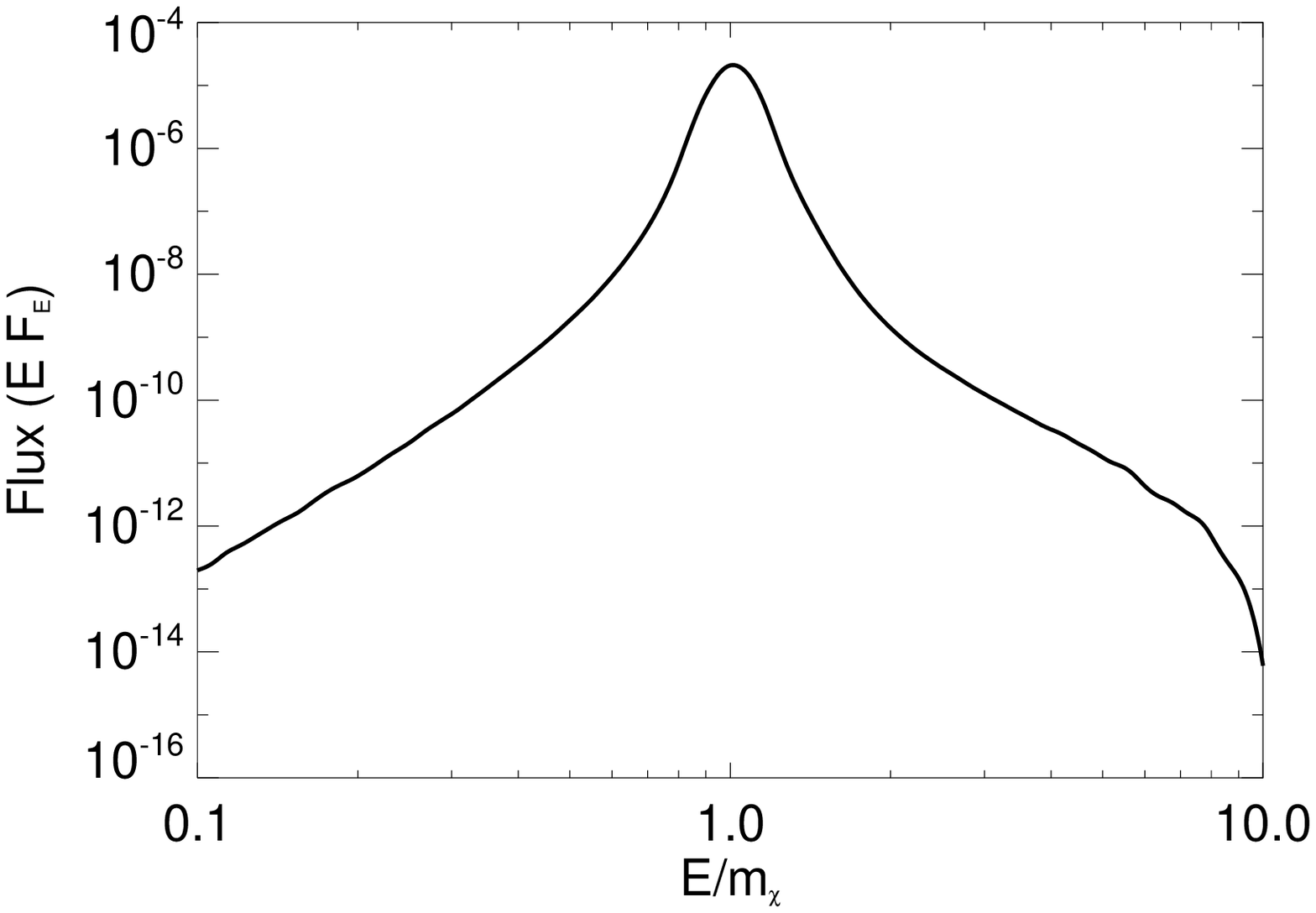}}
\end{center}
\end{figure}

As in \citet{Schnittman2013}, for the photons that reach infinity,
{\tt Pandurata}
can generate an image and spectrum of the
emission region. An example in shown in Figure \ref{fig:image_100} for
the annihilation signal from the unbound population around an extremal
black hole, limiting the emission signal to the region $r<100M$.
While the flux clearly increases towards the
center of the image, because the density and velocity profiles are
relatively shallow (see Fig.\ \ref{fig:dn_dr} above), the net flux is
actually dominated by emission from large radii. These annihilation
events are not very relativistic, so produce a strong, narrow peak in the
observed spectrum, centered at the DM rest mass energy. 

\begin{figure}[h]
\caption{\label{fig:image_2} Simulated image and of the annihilation
  signal around an extremal Kerr black hole, now considering only
  annihilations with $E_{\rm com} > 3m_\chi$. The
  observer is located in the equatorial plane with the spin axis
  pointing up. While the image appears off-centered, it is actually
  aligned with the coordinate origin. The photon energy $E$ is scaled
  to the dark matter rest mass $m_\chi$.}
\begin{center}
\scalebox{0.6}{\includegraphics*[55,450][360,720]{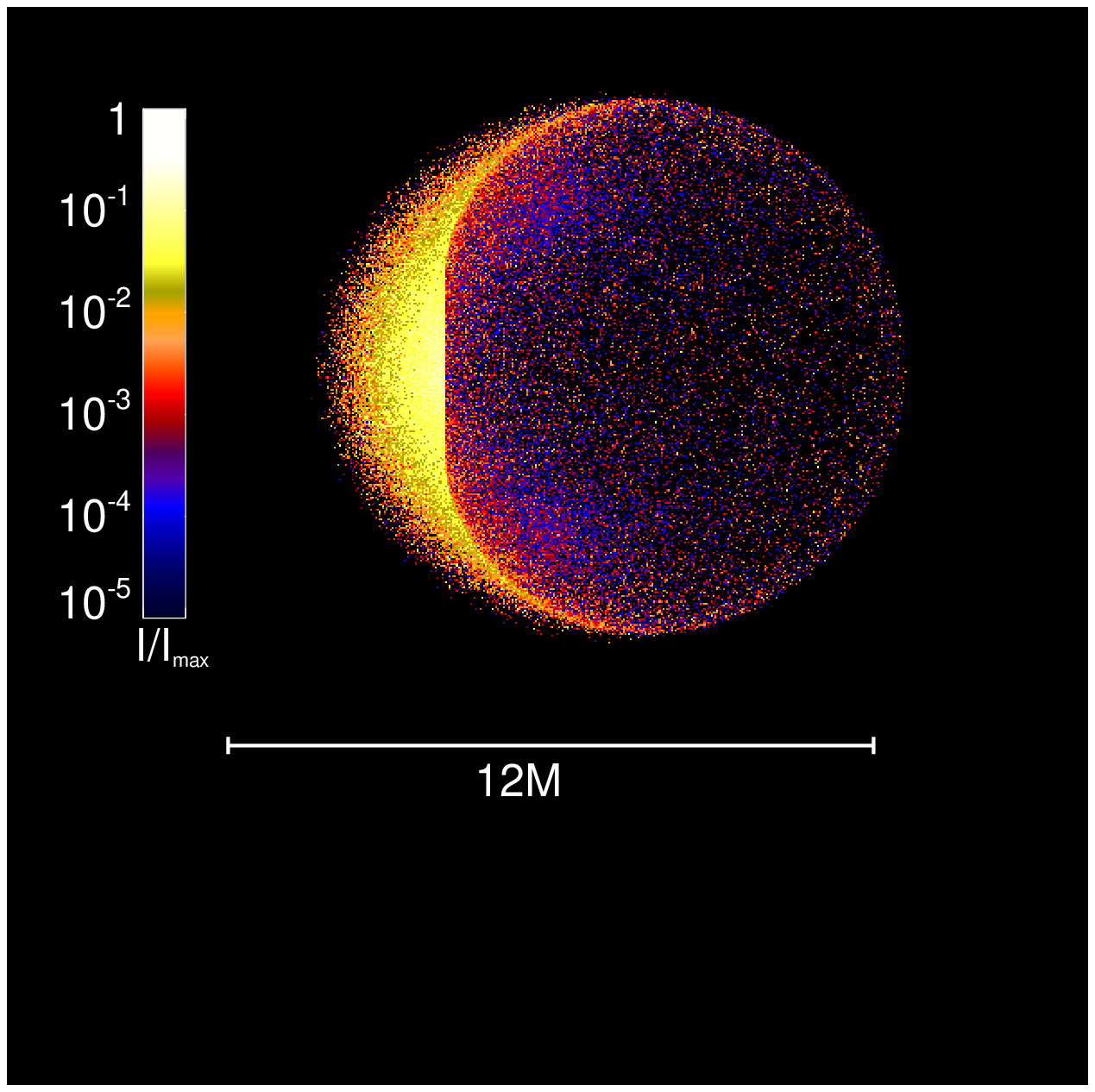}}
\scalebox{0.45}{\includegraphics*{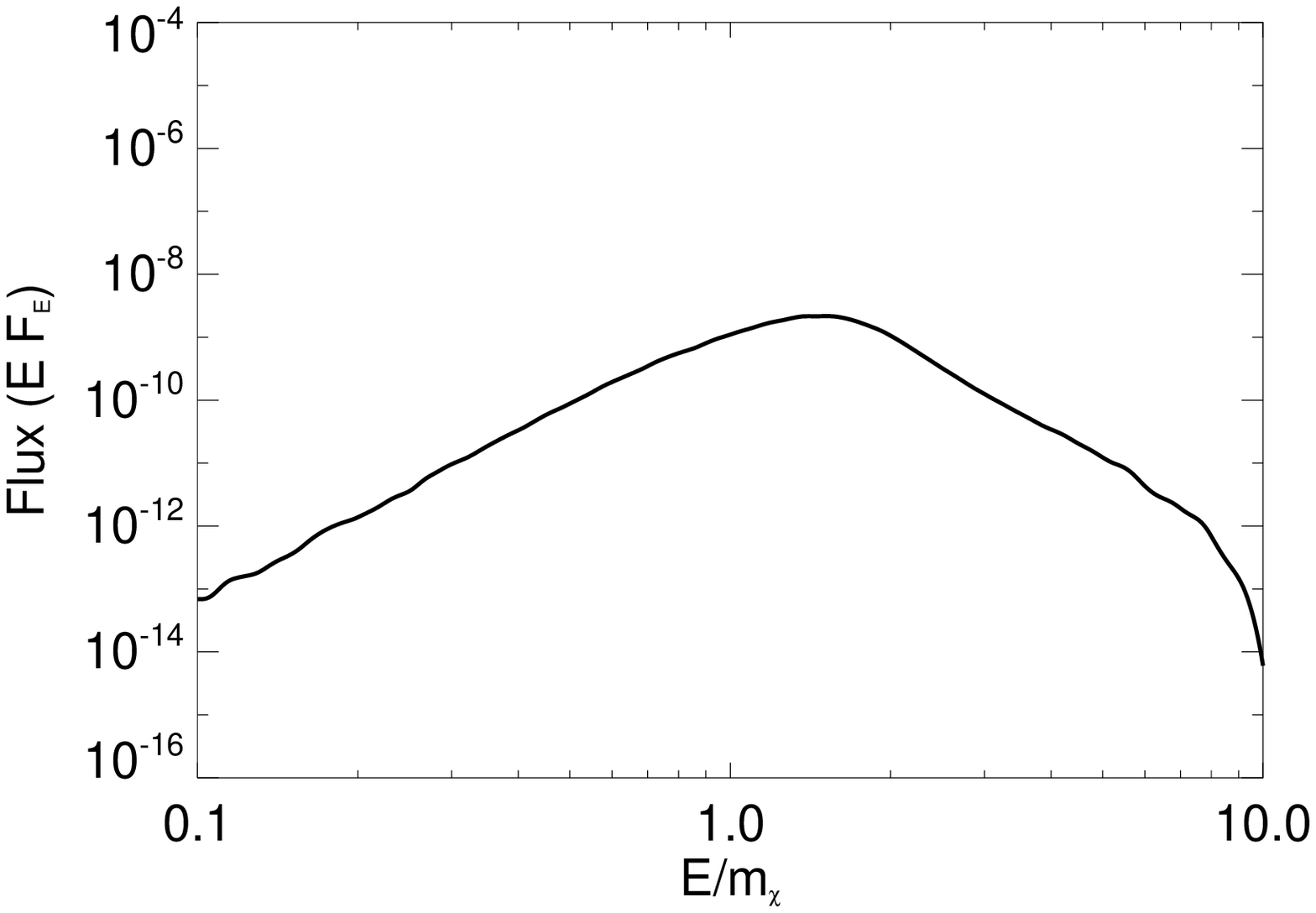}}
\end{center}
\end{figure}

The annihilation events occurring closer to the horizon
sample a much more energetic population of particles. Restricting ourselves
to only those events where the center-of-mass energy is greater than
$1.5\times$ the combined rest mass of the annihilating particles, we can zoom in to
the center of Figure \ref{fig:image_100}.
The result is shown in Figure \ref{fig:image_2}, now focusing on the
inner region within $r<6M$. At these small radii, the effects of black hole
spin become much more evident. One such effect is the characteristic
shape of the Kerr shadow, defined by the impact parameter of critical
photon orbits \citep{Chandra1983}. The observed flux is clearly
asymmetric, as the prograde photons originating from the left side of
the image have a much greater chance of escaping the ergosphere and
reaching a distant observer. 

There is another interesting feature of Figure \ref{fig:image_2} that
we believe is novel to this work. Namely, the purple lobes emerging
from the ``mid-latitude'' regions near the center of the image. These
are regions of greater photon flux, albeit very highly
redshifted. Recall, this image is created by considering only
annihilations with moderately high center-of-mass energy. Near the
equatorial plane, extreme frame dragging ensures that the velocity
dispersion is highly anisotropic, with most of the DM particles and
their annihilation photons getting swept
along on prograde, equatorial orbits. Above and below the plane, the
DM distribution is more isotropic, leading to a more isotropic
distribution of outgoing photons. Yet if one goes two far off the
midplane, it becomes more difficult for the photons to escape. At the
mid-latitudes, there is just enough frame dragging for photons to
escape, yet not so much that they get deflected away from the
observer. 

The spectrum corresponding to this
image is also plotted in Figure \ref{fig:image_2}. 
Not surprisingly, the red and blue wings of the annihilation line
shown in Figure \ref{fig:image_100} come
from the most relativistic events. As pointed out by
\citet{Piran1977}, even reactions with very high center-of-mass
energies will typically lead to photons with low energies as measured
at infinity, thus explaining the red tail of the annihilation
spectrum. The high-energy tail above $E=2m_\chi$ is due exclusively to
Penrose-process reactions where one of the annihilation photons has
negative energy and gets captured by the black hole
\citep{Penrose1969,Piran1975}. 

Earlier analytic work predicted
that the maximum energy attainable from the collisional Penrose
process was $2.6m_\chi$ for particles falling from rest at infinity
\citep{Harada2012,Bejger2012}. Because our calculation is fully 
numerical, it was able to reveal previously
unknown trajectories leading to very high efficiencies with $E >
10m_\chi$, as seen in Figure \ref{fig:image_2}. Closer inspection
revealed that these high-energy photons are created when an infalling
retrograde particle collides with a outgoing prograde particle that
has just enough angular momentum to reflect off the centrifugal
barrier, providing the necessary energy and momentum for the
annihilation photon to escape the black hole \citep{Schnittman2014,Berti2014}. 

\begin{figure}[h]
\caption{\label{fig:spectrum_2inc} Observed annihilation spectrum for
  the unbound DM population, as a function of
  observer inclination angle, considering only annihilations with
  $E_{\rm com}> 3m_\chi$. The black hole spin is $a/M=1$.}
\begin{center}
\scalebox{0.45}{\includegraphics*{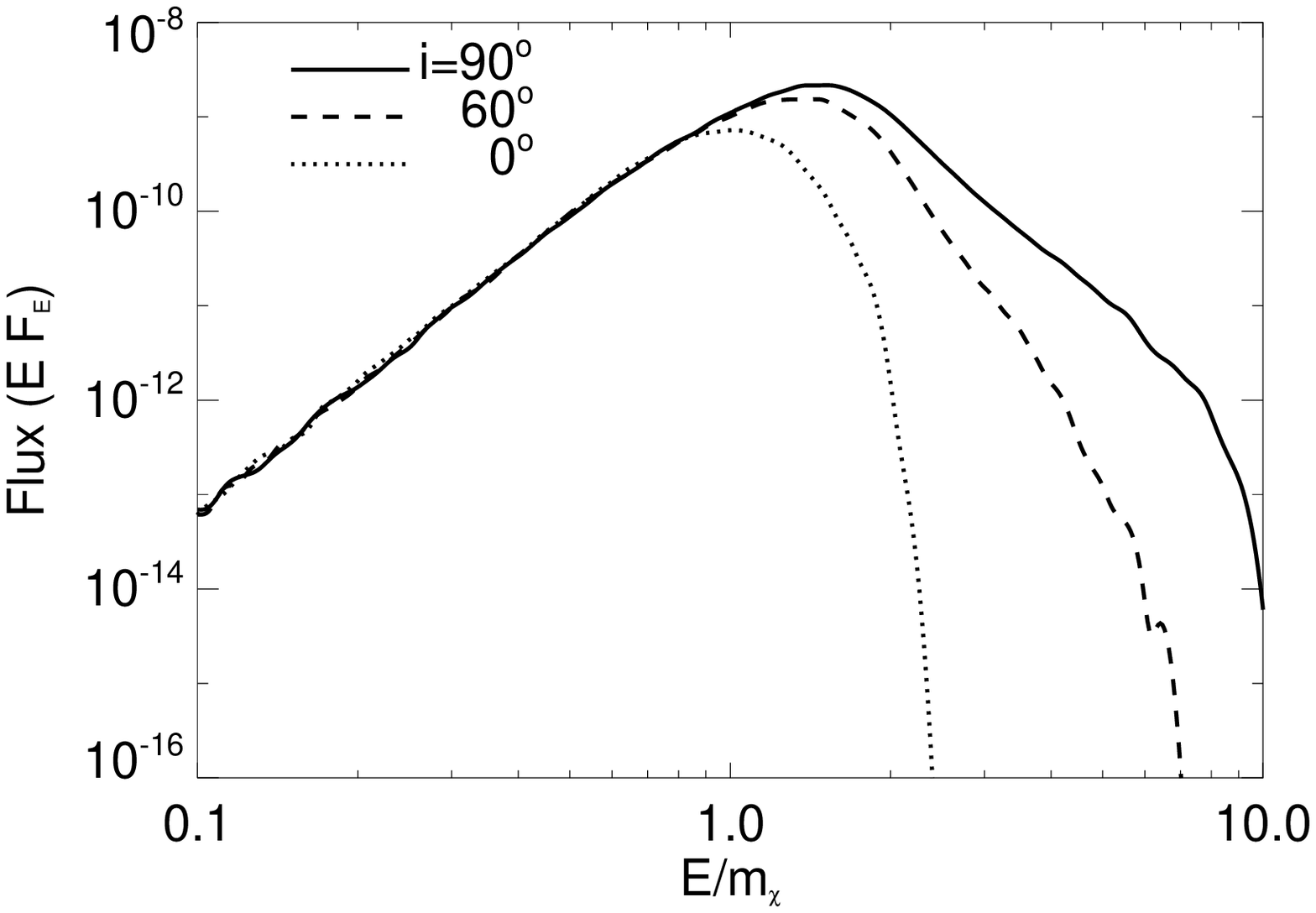}}
\end{center}
\end{figure}

Due to the strong forward-beaming effects within the ergosphere, the
escaping photon flux is highly anisotropic, with the peak flux and
highest-energy photons emitted in the equatorial plane. Figure
\ref{fig:spectrum_2inc} shows the predicted annihilation spectra for
observers at different inclination angles for the same DM profile as
shown in Figure \ref{fig:image_2}. Again, we restrict ourselves to the
highest-energy reactions with $E_{\rm com}>3m_\chi$.

\begin{figure}[h]
\caption{\label{fig:flux_rE} Flux reaching infinity (solid curves) and
  getting captured by the black hole (dashed curves), as a
  function of the center-of-mass energy and radius of
  annihilation, for both bound and unbound populations. The black
  hole spin is maximal. Note that the scale on the y-axis is
  arbitrary, and depends strongly on the annihilation cross sections
  and peak density. The radial flux profile, on the other hand, is a
  robust result for these populations.}
\begin{center}
\scalebox{0.45}{\includegraphics*{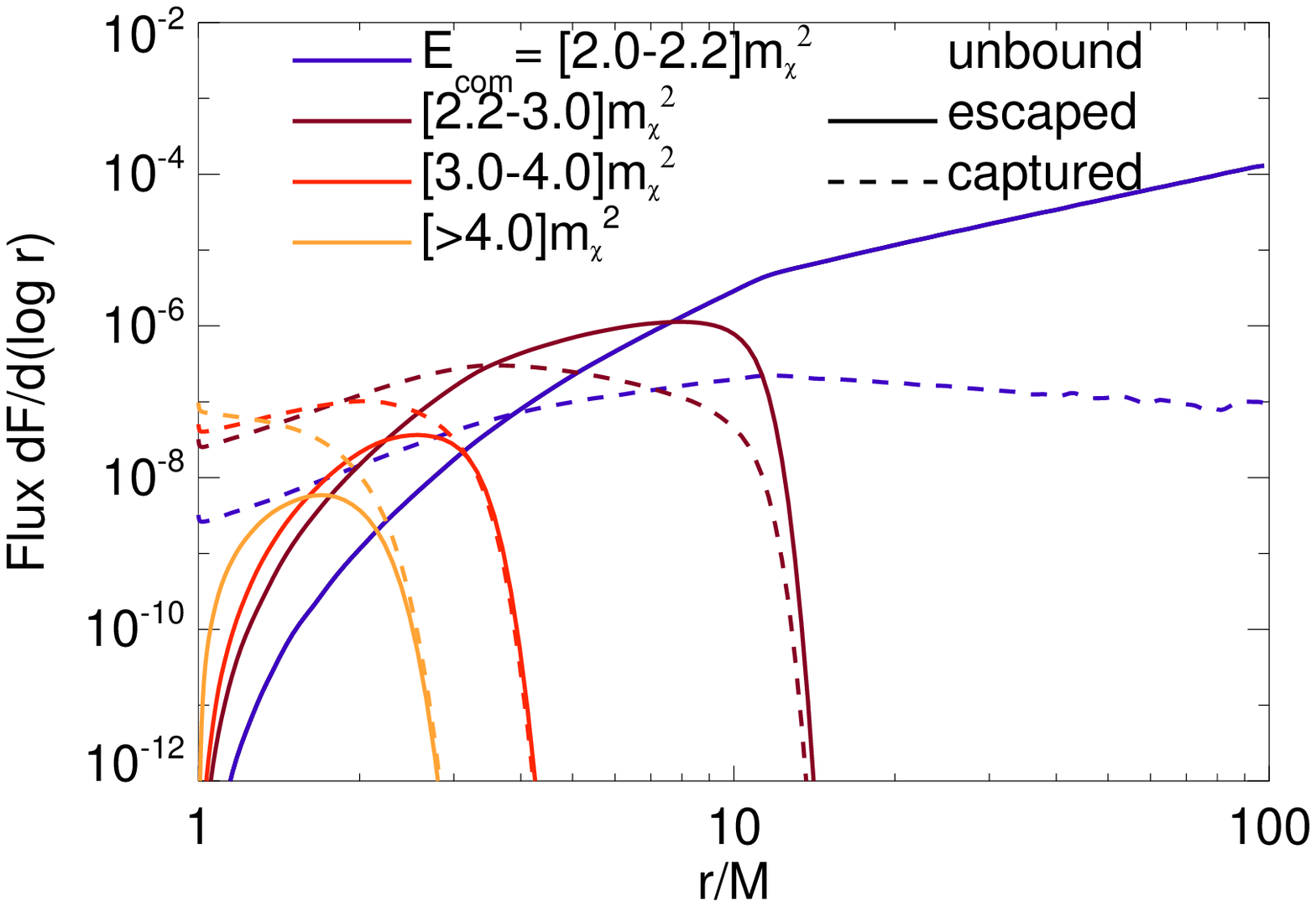}}
\scalebox{0.45}{\includegraphics*{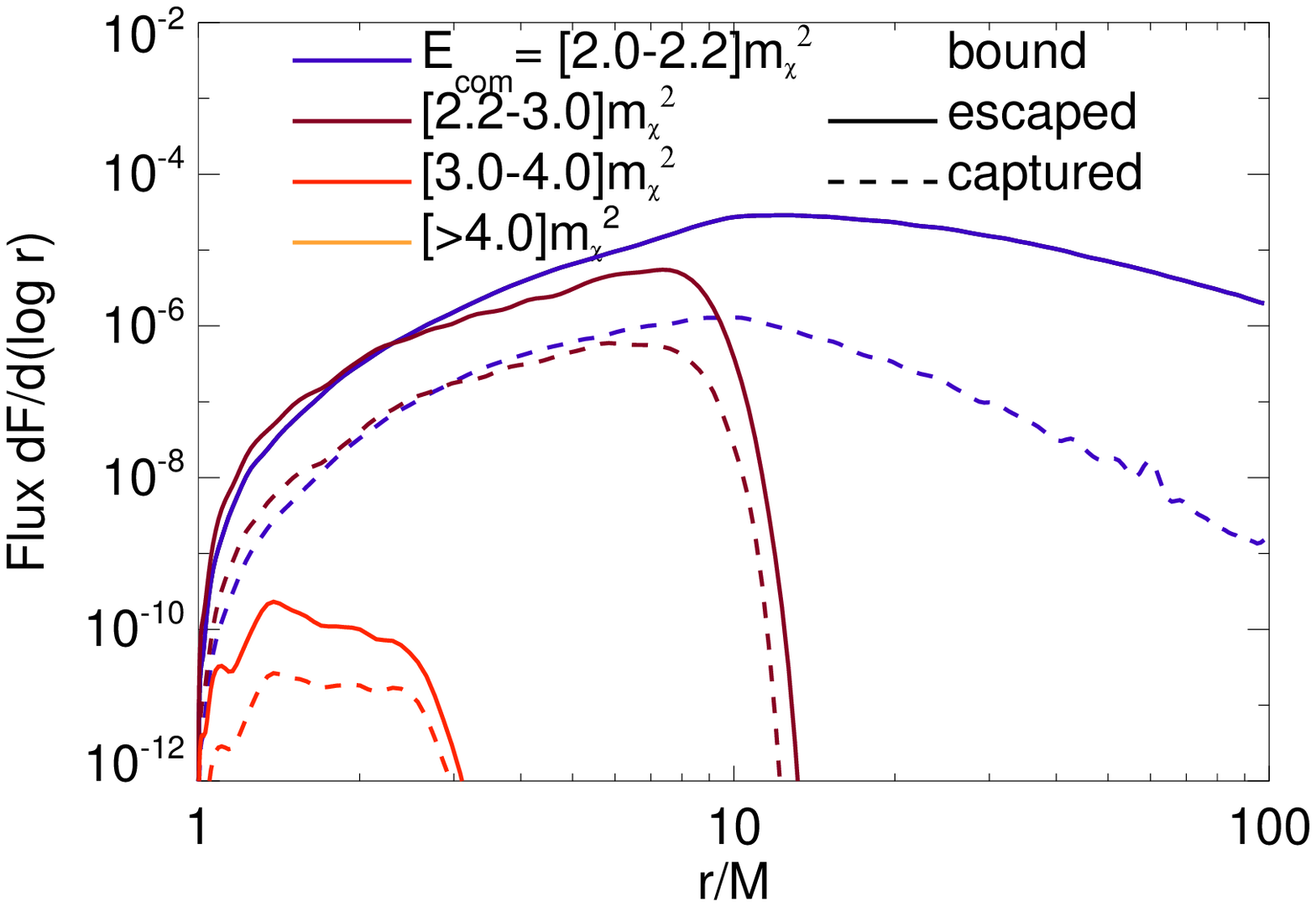}}
\end{center}
\end{figure}

It is also instructive to plot the annihilation flux as a function of
the emission radius. In Figure
\ref{fig:flux_rE} we show both the observed flux (solid curves) and
the flux that gets captured by the black hole (dashed curves) as a
function of radius, integrated over all observing angles. The emission
is further subdivided by the center-of-mass energy of the annihilating
particles. Of course, the photons emitted closer to the black hole have
a greater chance of getting captured. For the unbound population, the
total escape fraction 
ranges from $f_{\rm esc}=93\%$ at $r=10M$ down to $f_{\rm esc}(2M)=14\%$, and
$f_{\rm esc}(1.1M)=0.25\%$. At small radius, these numbers are somewhat smaller
than those calculated by \citet{Banados2011}, who only considered
critical trajectories in the equatorial plane, where the escape
probability is greatest. Yet at large radius, our distribution
includes particles with typically greater impact parameters, and thus
{\it greater} chance for escape. 

Another interesting feature of the curves in Figure \ref{fig:flux_rE}
is the very sharp cutoff above a critical radius for each energy
bin. This is a natural consequence of conservation of energy. Because
all unbound particles come in from rest at infinity with $E=m_\chi$,
the available kinetic energy in the center-of-mass frame is simply the
gravitational potential energy $Mm_\chi/r$ at that radius. For example, to reach a 
center-of-mass energy of $10\%$ above the rest mass energy, the
particles must fall within $r \approx 10M$. Also note that inside
$r\approx 4M$, most of the photons are captured, while outside of this
radius, most escape. This is in close agreement with what we found for
plunging orbits inside of the ISCO of a Schwarzschild accretion flow
in \citet{Schnittman2013a}. 

On the other hand, for the bound population of DM particles
(Fig.\ \ref{fig:flux_rE}b), which by
definition are {\it not} plunging, we find that the photon escape
fraction is more than $90\%$ at all radii, greatly increasing the
relativistic effects observable from infinity. This is consistent with
the classic calculation by \citet{Thorne1974} which found that for
thin accretion disks limited to circular, planar orbits outside the
ISCO, the fraction of emission ultimately captured by the black hole
was never more than a few percent, even for maximally spinning black
holes where the majority of the flux emerges from extremely close to
the horizon. 

\begin{figure}[h]
\caption{\label{fig:spectrum_spin} Observed spectrum as a function of
  black hole spin, for an observer at inclination $i=90^\circ$,
  considering only annihilations with $E_{\rm com}> 3m_\chi$.}
\begin{center}
\scalebox{0.45}{\includegraphics*{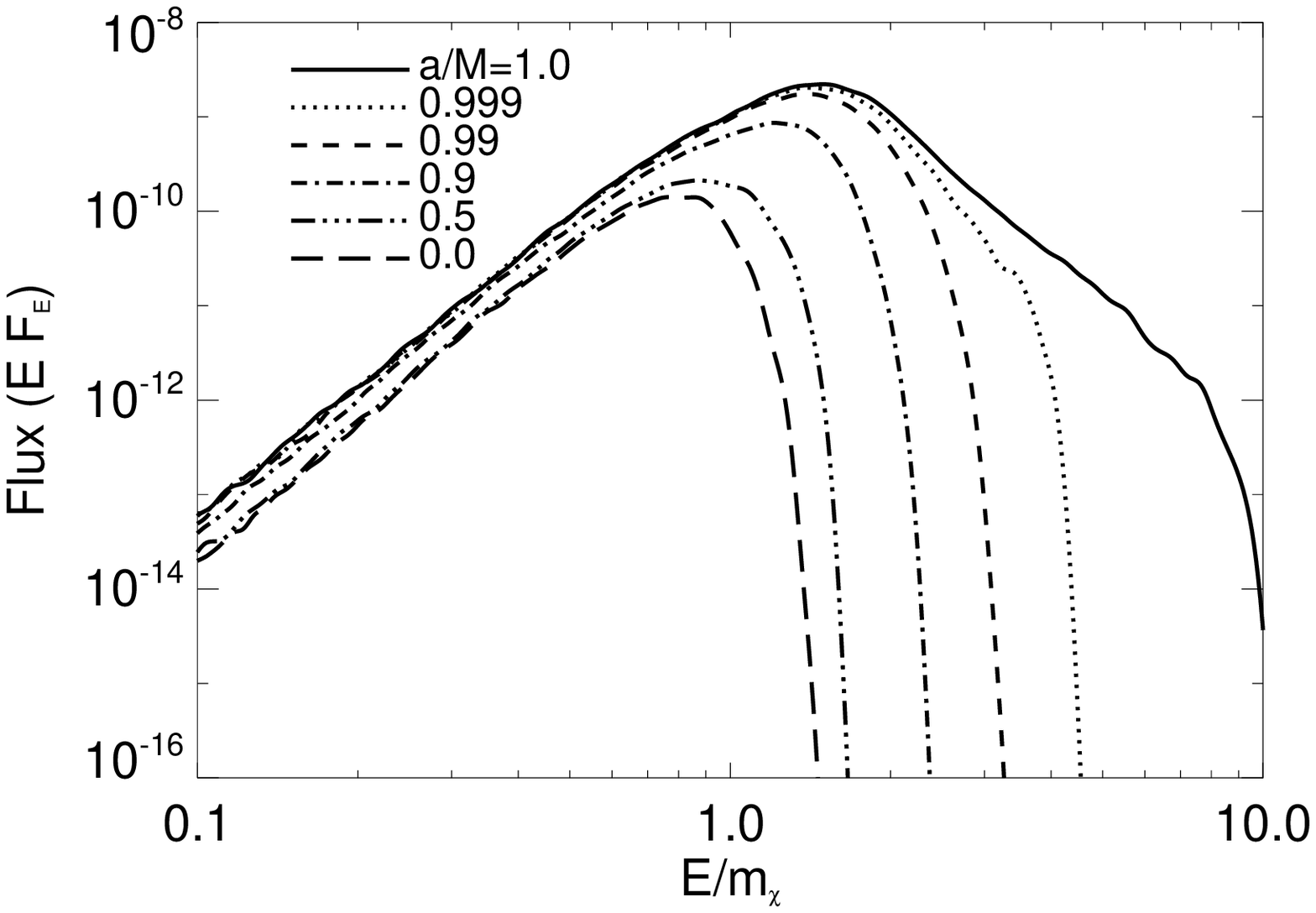}}
\end{center}
\end{figure}

As we showed in \citet{Schnittman2014}, the peak energy attainable
from particles falling in from infinity is a strong function of the
black hole spin. Now, considering the full phase-space distribution function of
the particles, we can see how the shape of the spectrum depends on
spin. In Figure \ref{fig:spectrum_spin} we plot the flux seen by an
equatorial observer, again limited to the high-energy annihilations
with $E_{\rm com} > 3m_\chi$. For even marginally sub-extremal spins,
the peak photon energy falls precipitously. As the spin decreases
further, the number of collisions with $E_{\rm com} > 3m_\chi$ also
decreases, thereby reducing the total flux observed. Lastly, the
decreasing spin also increases the critical impact parameter for
capturing prograde photons, making it harder for the annihilation flux
to escape to infinity. 

Recall from Section \ref{section:unbound} above that the density of
the unbound distribution scales like $n\sim r^{-1/2}$. From the
rate calculation in equation (\ref{eqn:rate}) we see that the
annihilation rate [events/s/cm$^3$] scales like $R(r) \sim
r^{-3/2}$. Including the volume factor $dV=4\pi r^2dr$ we can write
the differential annihilation rate as $dR/dr \sim r^{1/2}$. In other
words, the unbound contribution to the annihilation signal diverges at
large radius. In practice, the outer boundary can be set as the black
hole's influence radius, typically $10^{6-7} r_g$. This means that the
observed signal will essentially be a delta function in energy, with
only small perturbations from the relativistic contributions at small
$r$, and thus measuring spin from annihilation lines would be a very
challenging prospect indeed. 

Two possible effects provide a way around this problem, each with its
own additional uncertainties. One possibility is that the annihilation
cross section is a strong function of energy, increasing sharply above
some threshold energy. This is admittedly rather speculative, and in conflict
with leading DM models of self-annihilation \citep{Bertone2005}. On the other hand, we do
not even know what the dark matter particle is, or if there are many
DM species making up a rich ``dark sector,'' with all the beauty and
complexity of the standard model particles \citep{Zurek2014}. One
could easily imagine a DM analog of pion production via the collision of
high-energy protons, in which case the only reactions could occur
immediately surrounding a black hole, the ultimate gravitational
particle accelerator. In this case, by construction the annihilation
rate is dominated the region immediately surrounding the
black hole.

Another possibility is that the DM density is dominated by a
population of bound particles. As described above in section
\ref{section:bound}, this population arises
through the adiabatic growth of the black hole through accretion, capturing
marginally unbound particles while also making the bound particles
ever more tightly bound \citet{Gondolo1999,Sadeghian2013}. This
process will generally lead to a much 
steeper density profile, such as the $n\sim r^{-2}$ distribution we use
here. In this case, the differential reaction rate scales like $dR/dr
\sim r^{-5/2}$ so the annihilation spectrum is now dominated by the
particles at smallest radii. In both cases---energy-dependent cross
sections and a large bound population---the relativistic effects
described in Section \ref{section:unbound} (expanded proper volume and
time dilation) push the most important interaction region to even
smaller radii, and thus the annihilation spectra are even more
sensitive to the black hole spin. 

\begin{figure}[h]
\caption{\label{fig:spectrum_spin1000} Comparison of annihilation
  spectra from bound and unbound populations, including all emission
  out to $r=1000M$. The peak of the unbound signal will actually be
  even narrower, as it is dominated by annihilations at large radii
  with small relative velocities.}
\begin{center}
\scalebox{0.45}{\includegraphics*{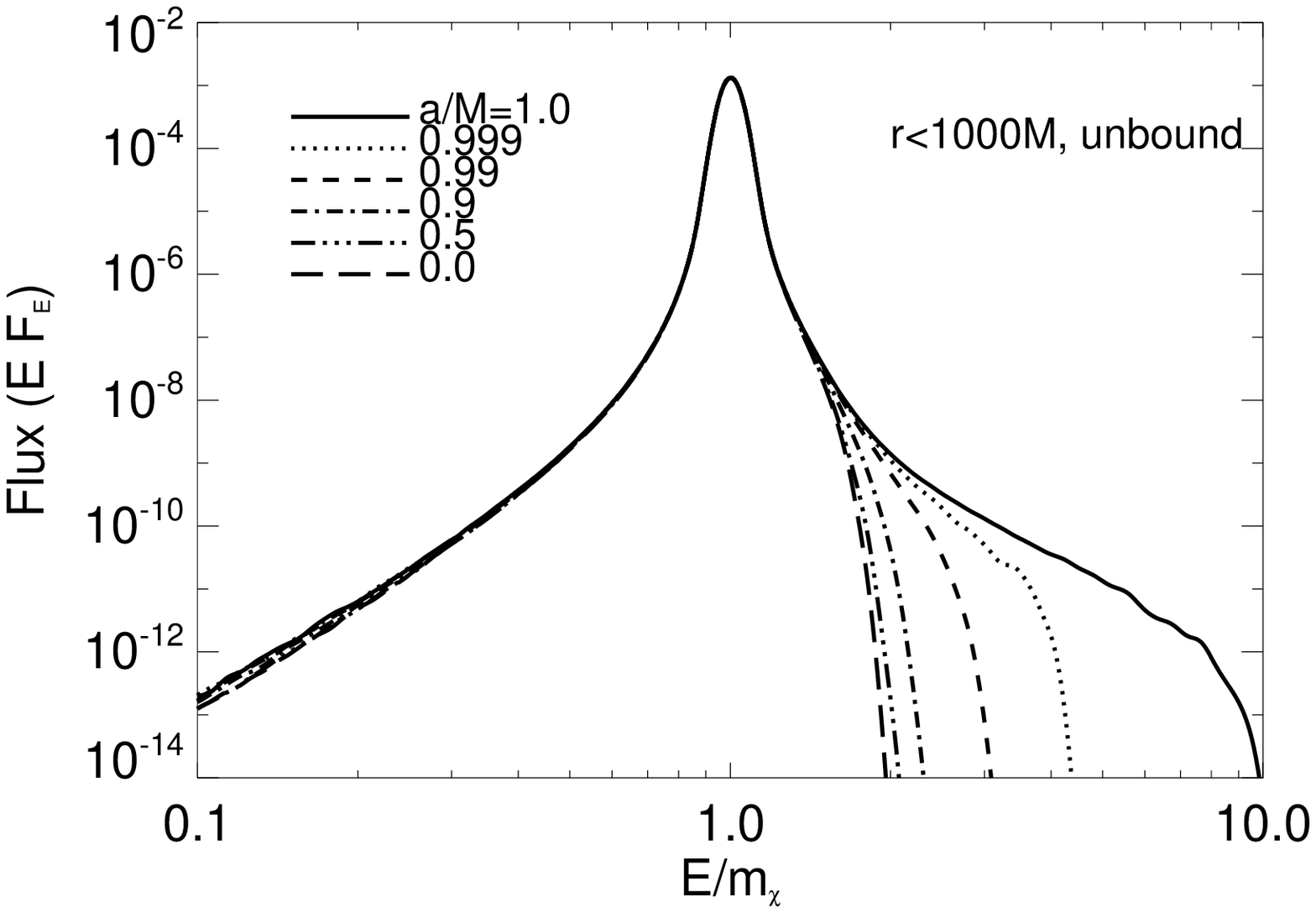}}
\scalebox{0.45}{\includegraphics*{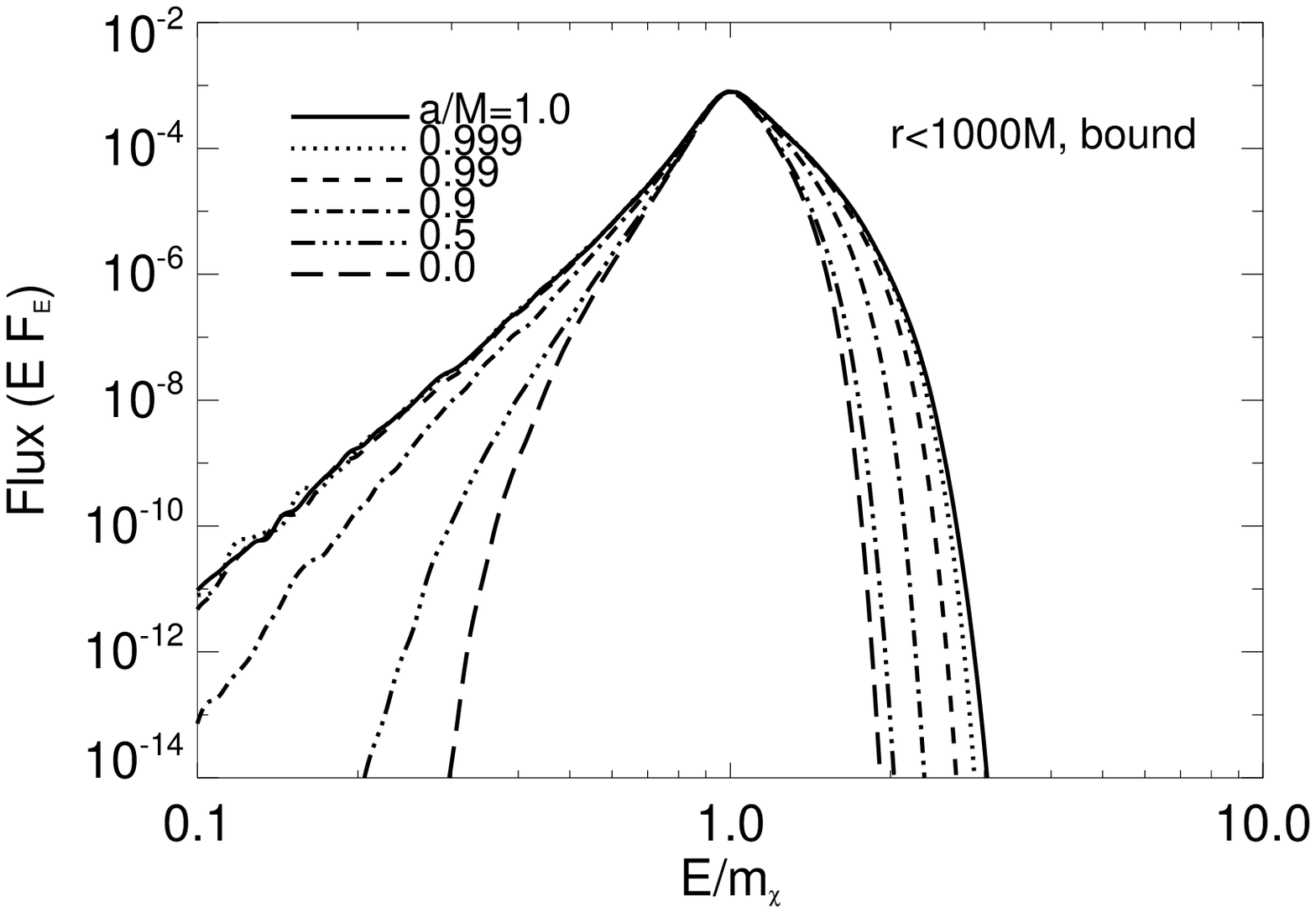}}
\end{center}
\end{figure}

In Figure \ref{fig:spectrum_spin1000} we show the annihilation spectra
for both the bound and unbound populations for a variety of spins, now
including emission out to $r=1000M$. The
relative amplitudes are somewhat arbitrary, because we don't know what
the relative densities of the two populations might be (see discussion
below in Sec.\ \ref{section:observability}), but it is almost certain
that the bound population should
dominate, possibly even by many orders of magnitude
\citep{Gondolo1999}. At the same time, the unbound signal will be even
narrower and have a greater amplitude peak than shown here, as it is
dominated by low-velocity particles at large radius. So while their
overall amplitudes are uncertain, the detailed shapes of the spectra
away from the central peak are relatively robust, depending only on
the properties of geodesic orbits near the black hole. 

In this broad part of the spectrum, the bound and unbound signals show
very different behavior. For non-spinning black holes, no particle can
remain on a bound orbit inside of $r=4M$ (see Fig.\ \ref{fig:n_rth}),
so there are no annihilation photons coming from just outside the
horizon, and these are the photons that produce the most strongly
redshifted tail of the spectrum. As the spin increases and the ISCO
moves to smaller and smaller radii, the line becomes steadily
broader. On the other hand, the unbound particles are found all the
way down to the horizon, where they can annihilate to highly redshifted
photons regardless of the black hole spin. 

Comparing Figures \ref{fig:df_2} and \ref{fig:df_2_bound}, we see that
the unbound particles probe a much greater volume of momentum space at
small radii. This in turn
leads to a greater chance of producing the extreme Penrose particles
that characterize the blue tail of the spectrum. Because all the bound
particles are essentially on the same prograde, equatorial orbits, it
is much more difficult to achieve annihilations with large
center-of-mass energies, so the high-energy cutoff in the spectrum is
much closer to the classical result for a single particle decaying
into two photons in the ergosphere \citep{Wald1974}. In short,
for bound particles the red tail of the spectrum is a better
probe of black hole spin, while for the unbound population, the blue
tail is the more sensitive feature. But in both cases, higher spin
leads to a broader annihilation line. 

\section{OBSERVABILITY}\label{section:observability}

In addition to the dependence on the dark matter density profile, the
amplitude of the annihilation spectrum will also depend on the unknown
dark matter mass and annihilation cross section. At this point, it is
only possible to use existing observations to set upper limits on these
unknown parameters. One major obstacle that has plagued nearly all
observational efforts to detect dark matter annihilation is the
existence of more conventional astrophysical objects such as active
galactic nuclei (AGN), pulsars, and supernova 
remnants, all of which are powerful sources of high energy gamma
rays. One solution to this problem is to focus on nearby dwarf
galaxies, which are thought to have a high DM fraction and are not
typically contaminated by AGN activity or significant star formation
\citep{Fermi2011}  (note, however, the recent work by \citet{GM2014},
which focuses on the contribution of black holes in dwarf galaxies). 

Yet for our purposes, it turns out that the strongest upper limits
actually come from the most massive galaxies with the most massive
central black holes. Massive elliptical galaxies have the added
advantage of being relatively quiescent both in nuclear activity and
star formation [e.g., \citet{Schawinski2007}]. As mentioned above, the
annihilation signal from 
the unbound population will be dominated by flux at large radius. It
is difficult enough to spatially resolve even nearby black holes'
influence radii with HST, much less gamma-ray telescopes, so any
potential annihilation signal will tell us little about the black hole
itself. 

Prospects for detection of an unambiguous black hole signature improve
if we consider annihilation models that include an energy dependence
to the dark matter cross section. For example, p-wave annihilation mechanisms will
have cross sections proportional to the relative velocity between the
two annihilating particles [see \citet{Chen2013,Ferrer2013} and references
therein]. Unfortunately, from equation (\ref{eqn:sigma2_r}) we see
that this would only lead to an additional factor of 
$r^{-1/2}$ in the integrand of equation (\ref{eqn:rate}), which would
still be dominated by the contributions from large $r$. 

\begin{figure}[h]
\caption{\label{fig:spectrum_sigma10000} Comparison of annihilation
  spectra from unbound populations, for two simple models of the dark
  matter cross section. All spectra are normalized to their peak
  intensity. For this comparison, all emission within $r=10^4M$ is
  included.}
\begin{center}
\scalebox{0.45}{\includegraphics*{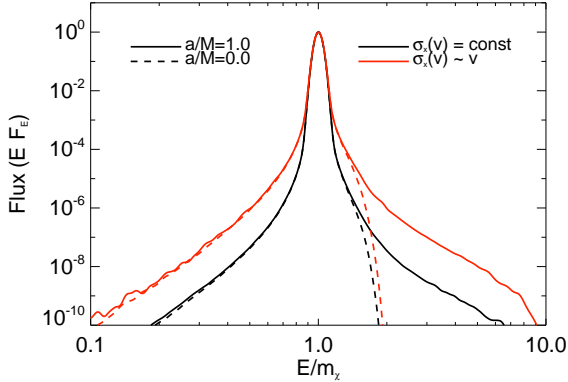}}
\end{center}
\end{figure}

This effect is shown in
Figure \ref{fig:spectrum_sigma10000}, which plots the predicted
spectra for two annihilation models: $\sigma_\chi(v) = \mbox{const}$
(black curves) and $\sigma_\chi(v) \propto v$ (red curves). The black hole
spins considered are $a/M=0$ (dashed curves) and $a/M=1$ (solid
curves), and in all cases only the unbound population is
included. Integrating out to $r=10^4 M$, we see only a slight
difference in the shape of the spectrum, with the $\sigma_\chi(v) \propto v$
model leading to a slightly broader peak (all curves are normalized to
give a peak amplitude of unity). 

Another possible annihilation model is based on a resonant reaction at
some energy above the DM rest mass, as suggested in
\citet{Baushev2009}. If the cross section increases sharply around a
given center-of-mass energy, this would have the effect of focusing in
on a relatively narrow volume of physical space around the black hole,
as in Figure \ref{fig:flux_rE}. 

Alternatively, the cross section could abruptly increase above a
certain threshold energy, if new particles in the dark sector become
energetically allowed, analogous to pion production via proton
scattering. In either the resonant or threshold models for the
annihilation cross section, one might imagine a pair of heavier,
intermediate dark particles getting created and then annihilating to
two photons as in the direct annihilation model. If, for example, the
mass of these intermediate particles is $1.5m_\chi$, then the observed
spectrum would look like those plotted in Figures \ref{fig:image_2}
and \ref{fig:spectrum_spin}. With a significant increase in the cross
section above such an energy threshold, these
relativistically-broadened spectra could in fact dominate over the
narrow line component produced by the rest of the galaxy.  

A less exotic option would be the simple density enhancement due to
the bound population. If this is sufficiently large, it would easily
dominate over the rest of the galaxy and also produce a
characteristically broadened line sensitive to both black hole spin
magnitude and orientation relative to the observer. Somewhat
ironically, one of the things that could ultimately limit the strength
of the annihilation signal from bound dark matter is annihilation
itself. If the adiabatic black hole growth occurred at high redshift,
then in the subsequent $\sim 10^{10}$ years, the bound population will
get depleted via self-annihilation at an accelerated pace due to its
high density \citep{Gondolo1999,GM2014}. 

On the other hand, if the black hole
grows through mergers, or experiences even a single merger since the
last extended accretion episode, it is quite likely that the bound
dark matter population could get completely disrupted. The details of
such an event are beyond the scope of this paper, but could be modeled
by following test particles bound to
each black hole through the merger, via post-Newtonian calculations
\citep{Schnittman2010} or numerical relativity \citep{vanMeter2010}. 

The observational challenge is readily apparent: the black holes with
the largest bound populations will tend to be in gas-rich galaxies
with a lot of accretion and high-energy nuclear activity that could
overwhelm the DM annihilation signal. The more
massive black holes, residing in gas-poor quiescent galaxies, are also
more likely to have lost their cloud of bound dark matter through a
history of mergers. Even in the event that a gas-rich spiral galaxy
hosts a quiescent nucleus, the black holes in those galaxies tend to
have lower masses \citep{Kormendy2013}. 

While the relation between black hole mass and dark matter density is
quite complicated for the bound population, it is relatively
straightforward to calculate for the unbound population, which we can
take as a lower bound on the DM density. Recall the
influence radius $r_{\rm infl}$ is the distance within which the 
gravitational potential is dominated by the black hole,
as opposed to the nuclear star cluster or dark matter halo. 
From equation (\ref{eqn:r_infl}) we see that the influence volume
scales like $r_{\rm infl}^3\sim M^{3/2}$, while the total mass
enclosed is---by definition---of the order
of $M$. If the dark matter and baryonic matter have similar profiles
(by no means a certainty!), then more massive black holes should have
lower surrounding DM density, with $n_{\rm infl }\sim M^{-1/2}$.  

Because the unbound DM density falls off more rapidly
outside the central core, the annihilation flux $F_{\rm unbound}$ will be dominated by
the contribution from around $r_{\rm infl}$, so we can estimate
\begin{equation}\label{eqn:F_standard}
F_{\rm unbound} \approx \frac{1}{D^2}n_{\rm infl}^2 \sigma_\chi v_{\rm infl} r_{\rm infl}^3
\sim M^{3/4}\, D^{-2}
\, ,
\end{equation}
with $D$ the distance to the black hole and the mean velocity at the
influence radius $v_{\rm infl}=\sigma_0$. 

If we consider a threshold energy annihilation model where all the
flux comes from inside a critical radius $r_{\rm crit} \sim
\mbox{few}\times r_g$, then
the density scales like $n_{\rm crit} \sim n_{\rm infl}(r_{\rm infl}/r_{\rm
  crit})^{1/2} \sim M^{-3/4}$ while the relative velocity scales like
$v_{\rm crit} \sim \sigma_0(r_{\rm infl}/r_{\rm crit})^{1/2} \sim
M^0$. The net flux then scales like
\begin{equation}\label{eqn:F_crit}
F_{\rm unbound} \approx \frac{1}{D^2} n_{\rm crit}^2 \sigma_\chi v_{\rm crit} r_{\rm
  crit}^3 \sim M^{3/2}\, D^{-2}
\, .
\end{equation}
In both cases, it appears that the brightest sources will be the
closest, as opposed to the most massive. 

Now consider the case where the annihilation signal is dominated by the
bound contribution, the bound density is in turn limited by a
self-annihilation ceiling as in \citet{Gondolo1999}, and there is a
threshold energy above which the cross section greatly increases. In
this case, the flux is simply proportional to the total volume within
the critical radius, so $F_{\rm bound} \sim M^3\, D^{-2}$. With this scaling, the
greatest flux will actually come from more distant, more massive black
holes. For example, NGC 1277, with a mass of $1.7\times 10^{10}
M_\odot$ and at a distance of 20 Mpc \citep{vandenBosch2012}, could
give an observed flux over a thousand times greater than our own Sgr
A$^\ast$!

Recent works by \citet{Fields2014} and \citet{GM2014} have argued
that current
Fermi limits of gamma-ray flux from Sgr A$^\ast$ and nearby dwarf
galaxies with massive black holes already place the strongest
limits on annihilation from DM density spikes. Based on the arguments
above, we believe that even stronger limits should come from more
distant, massive galaxies. The other important advance presented in
the present work is that, for either the energy-dependent cross sections,
or the steep density spikes, the annihilation signal will be
dominated by the region closest to the black hole, and thus a
fully numerical, relativistic rate calculation is absolutely
essential.

Lastly, we should mention that gamma-rays, while the primary
observable feature explored in this work, are not the only promising
annihilation product. High-energy neutrinos could also be produced
in some annihilation channels, particularly those with
energy-dependent cross sections like p-wave annihilation
\citep{Bertone2005}. While neutrinos obviously present many new
detection challenges, the successful commissioning of new
astronomical observatories like IceCube make this approach an
exciting prospect \citep{IceCube2013}. Furthermore, the non-DM backgrounds may
contribute significantly less confusion in the neutrino sky.

\section{DISCUSSION}\label{section:discussion}
As apparent in the previous section, there are still far too many
unknown model parameters to allow for quantitative predictions of the
annihilation flux from dark matter around black
holes. \citet{Sadeghian2013} put it best: ``There are uncertainties in
all aspects of these models. However one thing is certain: if the
central black hole Sgr A$^*$ is a rotating Kerr black hole and if
general relativity is correct, its external geometry is precisely
known. It therefore makes sense to make use of this certainty as much
as possible.'' We have attempted to follow their advice to the best of
our ability. 

Thus, in order of decreasing confidence, the results in this paper
can be summarized by the following:
\begin{itemize}
\item For a given DM density $n_{\rm infl}$ and velocity dispersion
  $\sigma_0$ at the black hole's influence radius, the fully
  relativistic, 5-dimensional phase-space distribution has been
  calculated exactly for any black hole spin parameter, covering the
  region from $r_{\rm infl}$ all the way down to the horizon.
\item Given this distribution function and a
  model for dark matter annihilation, the observed gamma-ray
  spectrum can be calculated by following photons from their creation
  until they are either captured by the black hole or reach the
  observer. Two important relativistic effects serve to increase the
  annihilation rate as compared to a purely Newtonian treatment: time
  dilation near the black hole effectively raises the density of the
  unbound population in a steady-state distribution being fed from
  infinity; and transforming from coordinate to proper distances
  greatly increases the interaction volume in the region immediately
  around the black hole (see Fig.\ \ref{fig:dV_dr}).
\item Our numerical approach has unveiled previously overlooked orbits
  that can produce annihilation photons with extreme energies, far
  exceeding previous estimates for the maximum efficiency of the
  collisional Penrose process \citep{Schnittman2014}. The peak energy
  attainable for escaping photons is a strong function of the black
  hole spin. 
\item The population of bound dark matter has also been calculated
  numerically, although this depends on two additional physical
  assumptions: a local isothermal velocity distribution with a
  virial-like temperature; and an overall radial power-law for the
  density, as found in \citet{Gondolo1999} and
  \citet{Sadeghian2013}. Including only the long-lived stable orbits,
  we found that the density peaks in the equatorial plane somewhat
  outside of the ISCO, forming a thick, co-rotating torus around the
  black hole spin axis. Because the bound population is not plunging
  towards the horizon, the emerging flux has a much greater chance of
  escaping the black hole.
\item The annihilation spectra from both the bound and unbound
  populations are sensitive to the spin parameter, but in opposite
  ways: the unbound spectrum varies mostly in the high-energy cutoff,
  with higher spins allowing higher-energy annihilation products; the
  bound population moves closer and closer to the horizon with
  increasing spin, giving a stronger red-shifted tail to the
  annihilation spectrum. Both bound and unbound
  spectra become more sensitive to observer inclination with
  increasing spin, as the spherical symmetry of the system is broken. 
\item For dark matter particle physics models with an energy-dependent
  cross section (particularly one that increases with
  center-of-mass energy), the annihilation spectrum will be a more
  sensitive probe of the black hole properties. For DM models incorporating
  a rich population of dark sector species, black holes may
  be the most promising way to accelerate these particles and observe
  their interactions.
\item The shape of the annihilation spectra is relatively robust, but
  the normalization is highly dependent on uncertain parameters such
  as the dark matter density profile and cross section. If the unbound
  density profile follows the baryonic matter, with the shallow slopes
  seen in core galaxies, 
  the observed flux should be a relatively weak function of black hole
  mass. If, on the other hand, the annihilation signal is produced by
  the most
  relativistic population within $r_{\rm crit} \sim \mbox{few}\times
  r_g$, then the signal could scale like $M^3$ and thus be dominated
  by the most massive black holes in the local Universe. 
\end{itemize}

While this paper has treated the bound and unbound particles
separately, future work will also consider the self-interaction
between these two populations \citep{Shapiro2014,Fields2014}, which
may lead to a single, self-consistent steady-state distribution with
density slope between $-1/2$ and $-2$.
Future work will also focus on developing a robust framework in which we
can use existing and future gamma-ray observations to constrain
various parameters of the particle physics (e.g., $m_\chi$,
$\sigma_\chi(E)$, and the annihilation mechanism, i.e., line vs
continuum) and astrophysical models ($n_{\rm infl}$, the bound
distribution normalization and slope, and the black hole mass, spin,
and inclination). While initial work will focus on setting upper
limits on reaction rates by looking at quiescent galaxies, our
ultimate ambition is nothing short of an unambiguous detection of dark
matter annihilation around supermassive black holes. 

\section*{Acknowledgments}

We thank Alessandra Buonanno, Francesc Ferrer, Ted
Jacobson, Henric Krawczynski, Tzvi Piran, Laleh Sadeghian, and Joe
Silk for helpful comments and discussion. Special gratitude is due to
HKB''H for providing us a world full of elegant wonder and beauty. 
This work was partially supported by NASA grants ATP12-0139 and
ATP13-0077. 

\newpage
\bibliography{dm.bib}

\end{document}